\documentclass[fleqn,usenatbib]{mnras}

\usepackage{newtxtext,newtxmath}

\usepackage[T1]{fontenc}

\DeclareRobustCommand{\VAN}[3]{#2}
\let\VANthebibliography\thebibliography
\def\thebibliography{\DeclareRobustCommand{\VAN}[3]{##3}\VANthebibliography}

\usepackage{graphicx}	
\usepackage{amsmath}	
\usepackage{bbm} 
\usepackage{verbatim}
\usepackage{subcaption}
\usepackage{url}
\usepackage{aas_macros}
\usepackage{siunitx} 
\usepackage{pdflscape} 

\defcitealias{2013MNRAS.430..174H}{Paper~1}
\defcitealias{2014MNRAS.443.1482H}{Paper~2}
\defcitealias{2016MNRAS.461.2025E}{Paper~3}
\defcitealias{2019MNRAS.490.5807E}{Paper~4}
\defcitealias{2023MNRAS.526.3421S}{Paper~5}

\title[Numerical modelling of radio galaxy lobes - Polarimetric simulations of realistic clusters]{Numerical modelling of the lobes of radio galaxies - Paper VI: Polarimetric simulations of Universal Pressure Profile cluster atmospheres}

\author[M. Stimpson et al.]{
M. Stimpson,$^{1}$\thanks{E-mail: mstimpson2@herts.ac.uk}
M.J. Hardcastle,$^{1}$
M. G. H. Krause$^{1}$
\\
$^{1}$Centre for Astrophysics Research, Department of Physics, Astronomy and Mathematics, University of Hertfordshire, College Lane, Hatfield, Hertfordshire AL10 9AB, UK\\
}

\date{Accepted XXX. Received YYY; in original form ZZZ}

\pubyear{2022}

\begin{document}
\label{firstpage}
\pagerange{\pageref{firstpage}--\pageref{lastpage}}
\maketitle

\begin{abstract}
We present the results of a polarization study based upon relativistic magnetohydrodynamic modelling of jets running into hydrostatic, spherically symmetric cluster atmospheres.  For the first time in a numerical simulation, we derive Faraday rotation measure maps (RM maps) from model cluster atmospheres based upon the Universal Pressure Profile (UPP), incorporating a temperature profile for a `typical' self-similar atmosphere described by only one parameter - $M_{500}$.  We compare our simulated polarization products with current observational data from VLA and LOFAR, as well as continuing investigations from our previous work, such as the detectability of the Laing-Garrington effect.  We also studied the variation of mean fractional polarization with cluster mass and jet power.  We produce simulated Stokes $Q$ and $U$ channel images and using the RM synthesis technique we create RM maps.  These data provide insight into what we should expect of current and future high-resolution polarimetric studies of AGN outflows as we were able to probe the limitations of the RM synthesis technique by comparing it with the RM map direct from our simulations.  Highlights of our study include clear reproduction of polarization enhancements towards the edges of radio lobes for suitable conditions and a demonstration that complex lobe morphologies with multiple emission and Faraday active regions interspersed as might be expected in some pole-on or perhaps precessing sources should be distinguishable in observations with current technology.  Given that the UPP is our most representative general cluster atmosphere, these numerical simulations, and the polarimetric properties derived from them, represent the most realistic yet for spherically symmetric atmospheres.
\end{abstract}





\section{Introduction}

\subsection{Faraday rotation}

Magnetized jet plasma inflates a cavity (sometimes referred to as a cocoon or lobe) which pushes back the surrounding plasma and cluster magnetic field.  It is the lobe which is responsible for polarized radio synchrotron emission, and as this emission passes through the surrounding magnetoionic medium it is subjected to Faraday rotation \citep{2004JKAS...37..337C,2004IJMPD..13.1549G} which causes the plane of polarization of the electromagnetic wave to be rotated by an angle due to the difference in the propagation speed of the two circularly polarized components.  The observed angle of polarization, $\psi_{\text{obs}}$, is a function of the square of the wavelength $\lambda^2$ and is related to the intrinsic plane of polarization, $\psi_{\text{int}}$ (i.e. that emitted at the source) by
\begin{equation}
\psi_{\text{obs}}=\psi_{\text{int}}+\text{RM}\,\lambda^2\label{FR}
\end{equation}
where RM is the Faraday rotation measure or simply `rotation measure'.  RM is proportional to the line-of-sight magnetic field ($B_{\parallel}$)and the electron density ($n_e$) \citep{1966MNRAS.133...67B}, expressed as
\begin{equation}
\text{RM} = \frac{e^3}{8\pi^2\epsilon_o m^2 c^3}\int n_e B_{\parallel}\,\mathrm{d}l,\label{RM}
\end{equation}
where  $\mathrm{d}l$ is the path length, $e$ is the electron change, $m$ is the electron mass, $c$ is the speed of light and $\epsilon_o$ is the permittivity of free space. RM is measured in rad$\,$m$^{-2}$ and $e^3/8\pi^2\epsilon_o m^2 c^3$ is a constant which in physical units has a value of $C = \num{2.62e-13}$ T$^{-1}$. The use of Faraday rotation to establish the magnetic field strength is limited by the distribution of polarized radio sources and the current sensitivity of radio telescopes \citep{2009ApJ...702.1230T, 2019MNRAS.490.4841L}.  Typical values for RMs of AGN jets fall in the range $\sim 1 - 10^4$ rad$\,$m$^{-2}$ with values in the core sometimes reaching $10^5$ rad$\,$m$^{-2}$ \citep{2010ApJ...725..750B}. Average values across the lobes tend to be much lower, \cite{1997A&A...328...12P} found values of $\sim 50$ rad$\;$m$^{-2}$ across the lobes of Pictor A; \cite{2024MNRAS.529.1626A} found that values were higher and more varied for the eastern lobe with peak values approaching twice the average (the eastern lobe is at a greater distance from us).  In comparision, Hydra A displays extremely large values for RM of between $-12,000$ rad$\;$m$^{-2}$ and $5000$ rad$\;$m$^{-2}$ \citep{2023ApJ...955...16B}.  Such variations between sources is mainly due to the impact of the AGN environment: for example Pictor A is situated in a poor cluster whereas Hydra A is in a relatively rich cluster.\\

As the lobe inflates, there is very little mixing of the ICM plasma with the synchrotron-emitting lobe-plasma and so the Faraday effects are often considered to be entirely external.  The material between the cocoon contact surface to the domain boundary is called the \textit{Faraday screen}, or the  \textit{rotation measure integration volume}; the exact volume will depend upon the shape of the cocoon and the viewing angle.  The introduction of a jet into the surrounding plasma will increase the rotation measure; as the cocoon expands, the plasma between the cocoon and the bow shock is compressed perpendicularly to the jet direction and expands toroidally, which results in an increase in magnetic field strength, particularly tangentially to the lobe's working surface, as demonstrated in the simulations of \cite{2011MNRAS.418.1621H} which identified such RM enhancements towards the edges of the visible lobe, which has also been reported in recent observations \citep{2022ApJ...937...45A}.  These findings were confirmed in the simulations of \cite{2024MNRAS.531.2532J} who also demonstrated that the edge enhancement effect is greater for higher power jets running into denser environments.  \cite{2024MNRAS.531.2532J} also investigated the two-dimensional fluctuations in RM by calculating the structure function \citep{1984ApJ...284..126S, 1996ApJ...458..194M} and concluded that the edge enhancement effect is a minor one and that a much greater influence on the RM is the shape of the jet cocoon, the `RM window'.\\

Rotation measures can be derived from multi-frequency observations of sources within or behind clusters by measuring the angle of polarization as a function of $\lambda$ (equation \ref{FR}).  It is necessary to have at least three wavelengths in order to remove the $\pm n\pi$ ambiguity which comes about from the fact that the measured angle $\psi_{\text{obs}}$ lies in the range $0<\psi_{\text{obs}}<\pi\,$ radians so that a pair of measurements would not tell us how many half-rotations have taken place between source and observer.  The contribution from our own Galaxy is then subtracted and using measured values of $n_e$ along our line of sight, we can recover a measure of the ICM field (equation \ref{RM}).  In practice, the use of RM to constrain the ICM field is limited to a few nearby clusters with sufficient numbers of embedded or background objects which can be detected and from which measurements can be taken; the Coma cluster is the best constrained example, having 12 polarized sources detected in and behind it \citep{2010A&A...513A..30B,2013MNRAS.433.3208B}.\\

Analytical approaches to modelling RM have been used by a number of authors employing various levels of complexity \citep[e.g.][]{1991MNRAS.253..147T, 1990ApJ...355...29K, 2008LNP...740..143F}.  The last of these used a tangled magnetic field and radial gas density distribution varying with the isothermal $\beta$-model.  Real Faraday screens are a good deal more complicated than analytic models; for example, the Faraday rotation maps of Hydra A \citep{1993ApJ...416..554T} indicate that the northern and southern lobes have opposite RM values (i.e. positive and negative) suggesting a magnetic field variation on the scale of $50\,$kpc.  In addition, there are small-scale fluctuations on the size of $\sim\!10\,$kpc throughout both lobes; these findings indicate a multi-scale magnetic field.  For greater realism we must move away from analytic approaches to numerical models which more faithfully incorporate observations of the density and magnetic field distributions within cluster environments; in particular the ground-breaking work of \cite{2004A&A...424..429M} which has been developed and used in our current model (refer to \cite{2023MNRAS.526.3421S}).  In terms of observations, the high Faraday resolution of LOFAR \citep{2013A&A...556A...2V} has enabled RM studies at metre wavelengths and improved the resolution of observations, although the challenge is that longer wavelengths are more impacted by RM induced depolarization and so the sensitivity of observations needs to be greater.

\subsection{Polarization}

Optical and radio emission from the jets and lobes of Active Galactic Nuclei (AGN) is synchrotron, which is intrinsically highly polarized as seen for example in M87 \citep{1956ApJ...123..550B, 1991AJ....101.1632B}.  Theory indicates that, for optically thin emission and a uniform magnetic field, the polarization is nearly $70$\% \citep{1970ranp.book.....P} (see Appendix \ref{100_frac}).  In observations, however, the degree of polarization is often much less, for example \cite{1985ApJS...59..513A} found typical values of around $5$\%, and more recently \cite{2023MNRAS.519.5723O} found a median of $1.8$\% using data from the LOFAR Two-metre Sky Survey.  This is due to beam depolarization effects; along the line of sight, contributions from different regions with opposite orientations of magnetic field strength combine and cancel out and so reduce polarization.  In order to recover high polarization fractions we need to consider observations at high resolution and/or frequency.  Pictor A is a very bright discrete FRII radio source, well known for its round lobes.  Using data from the Very Large Array (VLA), \cite{1997A&A...328...12P} found polarization values of between 30 and 60 percent along the lobe edges and between 10 and 20 percent within the central regions of the lobes.  A recent study of Pictor A using the MeerKAT radio telescope \citep{2024MNRAS.529.1626A} also identified peak values of around 60 percent.  In comparison, observations of Hydra A reveal even greater variability in polarization across lobes, values ranging from as low as 2 percent in the inner regions to 75 percent at the lobe edges \citep{2023ApJ...955...16B}.  This wide range of values from various observations is expected as polarization depends upon the observation frequency, the properties of the source, the cluster environment and where in the source is being observed.\\

\subsection{Depolarization}

\cite{1966MNRAS.133...67B} showed that the radiation from sources of synchrotron radiation is depolarized as a result of Faraday rotation by the magnetised material which surrounds them.  Depolarization is the decrease in the percentage polarization that is caused by differential Faraday rotation between the source and observer (we refer to this as external depolarization, as opposed to internal depolarization which takes place within the lobes).  A fully resolved foreground Faraday screen  produces only Faraday rotation and not depolarization; but if we have finite resolution then beamwidth depolarization will result.  This occurs when the width of the observing beam is greater than the size of the fluctuations in the Faraday screen (as a result of variations in density and/or magnetic field) and so radiation with a similar position angle $\psi$ but opposite orientation will be averaged out and so the polarization reduced (i.e. depolarization will result).  In addition, bandwidth depolarization is caused by the variation of rotation angle across the frequencies being observed.   

External depolarization is frequency (or wavelength) dependent, the standard parametrization derived by \cite{1966MNRAS.133...67B} using the assumption of a Gaussian distribution of depths is (henceforth referred to as the `Burn law') is:
\begin{equation}\label{Burn_law}
p=p_i \exp{(-2C^2 ( n B_{\parallel} )^2_f d \ell \lambda^4 )}
\end{equation}
where $p$ is the observed fractional polarization and $p_i$ its intrinsic value, $C$ is a constant (see above), $( n B_{\parallel} )^2_f$ is the variance of the product of electron density and field strength along the line of sight, $\ell$ is the source size and $d$ is the characteristic size of field reversals, where we require that $d$ is significantly smaller than the physical size of the beam.  Clearly the problem here is the notion that there \textit{is} a characteristic scale that can be defined.  Observations show that a combination of resolved Faraday rotation and depolarization implicitly invalidates this simple assumption; in reality there is a power spectrum of values, which is taken into account in numerical models by giving the spectrum a Kolmogorov-like 3D turbulent slope \citep{2014MNRAS.443.1482H, 2016MNRAS.461.2025E, 2019MNRAS.490.5807E, 2023MNRAS.526.3421S}.

We can model depolarization by considering the thermal material in each cell between the lobe and observer (within the lobe); roughly speaking, this material rotates the plane of polarization by the angle
\begin{equation}
\phi = C n_{\text{th}} B_{\parallel} \lambda^2 \delta z
\end{equation}
where C is defined above, $n_{\text{th}}$ is the thermal density, $\lambda$ is the observation wavelength, $B_{\parallel}$ is the component of magnetic field along the line of sight (i.e. along the z-direction) and $\delta z$ is the length of the cell.  The polarization is expected to be a function of observing wavelength; In general we can expect depolarization effects to become important when $C n_{\text{th}} d B_{\parallel} \lambda^2 \ell \approx 1$ where $\ell$ is the distance through the region along the line of sight; from this we can calculate a critical frequency, below which depolarization will become important:
\begin{equation}\label{crit_freq}
\nu \approx c \sqrt{C n_{\text{th}} | d B_{\parallel} | \ell}.
\end{equation}

\subsection{Rotation measure synthesis}

The rotation measure is often taken to be the slope of the polarization angle $\psi$ versus $\lambda^2$ \citep[e.g.][]{1979A&A....78....1R}:
\begin{equation}
\text{RM} = \frac{\mathrm{d} \psi (\lambda^2)}{\mathrm{d} \lambda^2};
\end{equation}
however, if there is more than one emitter along the line of sight (LOS), then the slope will not be a constant and it can be a non-trivial task to disentangle the sources into separate components.  More advanced analysis uses the RM synthesis technique \citep{1966MNRAS.133...67B, 2005A&A...441.1217B, 2019MNRAS.490.4841L}.  Firstly we replace RM with the quantity $\phi$, the `Faraday depth'; a value related to the properties of the Faraday rotating plasma by the equation:
\begin{equation}\label{FR2}
\phi \propto \int_{\text{source}}^{\text{telescope}} n_e \mathbf{B} \cdot \mathrm{d} l
\end{equation}
(compare with equation \ref{RM}).  Using this concept, \cite{1966MNRAS.133...67B} introduced the Faraday dispersion function $F(\phi)$, which describes the intrinsic polarized flux as a function of Faraday depth.  By expressing the polarization vector as an exponential ($P=pe^{2i\psi}$), using equation \ref{FR2} and integrating over all Faraday depths, we obtain
\begin{equation}
P(\lambda^2) = \int^{+\infty}_{-\infty} F(\phi)e^{2i \phi \lambda^2} \mathrm{d}\phi,
\end{equation}
where $P(\lambda^2)$ is the (complex) observed polarization vector [$P(\lambda^2)=Q(\lambda^2)+iU(\lambda^2)$].  This relation takes the form of a Fourier transform and can be inverted to express the intrinsic polarization in terms of observable quantities as follows:
\begin{equation}
F(\phi) = \int^{+\infty}_{-\infty} P(\lambda^2)e^{-2i \phi \lambda^2} \mathrm{d} \lambda^2.
\end{equation}
We deal only with discrete and positive values of $\lambda^2$, and so the form of this equation which is used in practice is a discrete sum, which gives the reconstructed Faraday dispersion function
\begin{equation}
\tilde{F}(\phi) = K \sum^N_{n=1} W_n \tilde{p}_ne^{-2i\phi(\lambda_n^2-\lambda_0^2)},
\end{equation}
where $W_n$ are weights which can differ from unity, and $K$ is the inverse sum of the weights.  The term $\lambda^2_0$ has been included as \cite{2005A&A...441.1217B} demonstrated that if we give it a value equal to the weighted mean of the observed $\lambda_n^2$, then a better behaved response function results.  The reconstructed Faraday dispersion function can be expressed as:
\begin{equation}
\tilde{F}(\phi) = F(\phi) \ast R(\phi)
\end{equation}
where $\ast$ denotes convolution and $R(\phi)$ is the rotation measure spread function (RMSF), written in discrete form as:
\begin{equation}
R(\phi) = K \sum^N_{n=1} W_n e^{-2i\phi(\lambda_n^2-\lambda_0^2)}.
\end{equation}

RM synthesis is used to minimise the effects of $n\pi$ ambiguity, as well as to recover emission at multiple Faraday depths along a particular LOS.  The RM synthesis technique is widely used in the polarization analysis of observations such as those from LOFAR \citep[e.g.][]{ 2024MNRAS.527.9872G, 2023A&A...674A.119S, 2023MNRAS.518.2273C, 2022MNRAS.512..945C, 2021MNRAS.502..273M} and the Karl Jansky Very Large Array \citep[e.g.][]{2023A&A...675A..51D, 2023ApJ...955...16B, 2022A&A...666A...8S}.  We implement the RM synthesis technique using {\scriptsize{PYRMSYNTH}}\footnote{\url{https://github.com/mrbell/pyrmsynth}}, a Python script developed primarily for LOFAR Stokes $Q$ and $U$ cubes.

In this paper we aim to use the most realistic spherically symmetric cluster atmosphere to recover the polarization properties knowing the physical conditions of the clusters, which will give us insight into whether we can use the RM synthesis technique to recover the polarization properties of real cluster observations.  In Section 2 we present the UPP cluster atmosphere used in this study, our modelling of this and the numerical methods used to create our synthetic polarization quantities.  We present our simulated results in Section 3.  Our summary and conclusions are found in Section 4.

\section{Numerical simulation}

\subsection{The Universal Pressure Profile cluster atmosphere}\label{UPP_section}

Based upon observations of X-ray clusters with \textit{Chandra} and on numerical simulations on scales larger than these, \cite{2007ApJ...668....1N} built upon the work of \cite{1995MNRAS.275..720N} and proposed a `generalised Navarro, Frenk and White' (GNFW) model expressed in terms of the gas pressure of the cluster.  The version below is that presented by \cite{2010A&A...517A..92A} and is written in terms of the average scaled pressure $\mathbbm{p}$ at a normalized distance $x$ (equation \ref{UPP2}) from the cluster centre, the profile is
\begin{equation}
\mathbbm{p}(x)=\frac{P_0}{\left(c_{500}x\right)^{\gamma}\left[1+\left(c_{500}x\right)^\alpha\right]^{\left(\beta-\gamma\right)/\alpha}},\label{UPP}
\end{equation}
where $P_0$ is the pressure at the centre of the cluster and the parameters $\gamma,\alpha, \beta$ are respectively the central slope $\left(r \ll r_s\right)$, intermediate slope $\left(r \sim r_s \right)$ and outer slope $\left(r \gg r_s\right)$.  The scale radius $r_s$, is defined as the radius where the logarithmic slope of the density profile is $\alpha=-2$; and the concentration is defined as $c_{500}\equiv R_{500}/r_s$.  $R_{500}$ represents the radius of the cluster corresponding to a mean mass density contrast of $500$ times the critical density of the Universe.  These parametrized values are linked to real values of pressure $P(r)$ and radial distance $r$ using the scaling relations
\begin{equation}
P(r)=p(x)P_{500} \qquad \text{and} \qquad x \equiv r/R_{500}\label{UPP2}
\end{equation}
where $p(x)$ is the normalized pressure and is linked to the average scaled profile $\mathbbm{p}(x)$ by an empirical term which reflects the deviation from standard self-similar scaling:
\begin{equation}
p(x)=\mathbbm{p}(x) \left[ \frac{M_{500}}{\num{3e14} \text{h}^{-1}_{70}M_{\odot}} \right]^{\alpha(x)}\label{mass_dep}
\end{equation}
where $\alpha(x)$ is a variable in $x$ linked to the mass of the cluster and the dimensionless Hubble constant h$_{70}=$ h/H$_0$ where H$_0=70\,$kms$^{-1}$Mpc$^{-1}$.  $P_{500}$ is the `characteristic pressure' which is dependent upon mass and redshift as follows
\begin{equation}
P_{500}=\num{1.65e-3}h(z)^{8/3} \left[ \frac{M_{500}}{\num{3e14} \text{h}^{-1}_{70}M_{\odot}} \right]^{2/3} \text{h}^2_{70}\,\text{keV}\,\text{cm}^{-3}\label{UPP3}
\end{equation}
where $h(z)$ is the ratio of the Hubble constant at redshift $z$ to its present value, $h(z)=H(z)/H_0$.  $M_{500}$ is the mass contained within the radius $R_{500}$ at which the mean mass density is $500$ times that of the critical density of the Universe at the cluster redshift $\rho_c(z)$.  $M_{500}$ and $R_{500}$ can be found from one another; from the definition of $M_{500}$ we have
\begin{equation}
M_{500}=\frac{4\pi}{3}R^3_{500}500\rho_c(z)\qquad \text{where} \qquad \rho_c(z)=\frac{3H(z)^2}{8\pi G}\label{M500R500}
\end{equation}
where $G$ is the gravitational constant and $H(z)=H_0\sqrt{\Omega_M(1+z)^3+\Omega_{\Lambda}}$ where, for a flat $\Lambda$CDM cosmology, $\Omega_M=0.3$ and $\Omega_{\Lambda}=0.7$.  These relations are the GNFW model and the correct choice of parameters will result in a very good fit to the pressure profiles of galaxy clusters (as shown in Appendix C of \cite{2010A&A...517A..92A}).  From the UPP cluster pressure profile described here and using a cluster temperature profile we can recover the density profile using the ideal gas law $p=nkT$.

\cite{2010A&A...517A..92A} derived an average GNFW profile and for their choice of parameters the scaled pressure profiles do not show any significant dependence on mass; in other words equation \ref{mass_dep} reduces to $p(x)=\mathbbm{p}(x)$ and so their model is self-similar.  When X-ray measurements are used to estimate cluster masses, it is assumed that the cluster is in equilibrium (i.e. a perfectly relaxed cluster); however, observations indicate that clusters are not all relaxed as non-thermal pressure support is also present \cite[e.g.][]{2004A&A...426..387S, 2011MNRAS.410.1797S,  2015MNRAS.453.3699W, 2016Natur.535..117H, 2016A&A...585A.130H, 2019A&A...621A..40E, 2018ApJ...861...71S}; furthermore, simulations of clusters undergoing mergers or feedback processes substantiate these observations \citep[e.g][]{2012A&A...544A.103V, 2017MNRAS.464..210V, 2014ApJ...782..107N, 2017MNRAS.467.3737G, 2022MNRAS.514..313B} demonstrating that cluster formation leads to significant non-thermal gas processes such as turbulent flows and bulk motions.  Neglecting the kinematics leads to a systematic underestimation of the masses of galaxy clusters: this is the hydrostatic mass bias. \cite{2021ApJ...908...91H} describe how they employ a simulation (the Mock-X analysis framework) devised by \cite{2021MNRAS.506.2533B} which is able to model the evolution of clusters, including the non-thermal pressure support, and simulate X-ray emission.  Their study leads to the debiased values for the GNFW parameters (the values used in our model) which can be considered to be more accurate than those provided by previous studies; in addition, they confirmed the self-similarity conclusion of \cite{2010A&A...517A..92A} in that this set of generalised parameters does not depend upon mass.

In a \textit{cool core} cluster the temperature rises steeply away from the centre and reaches a peak of around a tenth of $R_{500}$, then reduces gradually towards large radii.  The core is believed to be at a lower temperature as a result of radiative cooling; the inner regions are at a higher pressure than the $\beta$-model predicts and this leads to a greater luminosity and so shorter cooling time than the rest of the cluster.  Without compensation through a heating mechanism, the core temperature falls \citep{1994ARA&A..32..277F}.  In this study we employ the \textit{cool core} temperature profile described by \cite{2006ApJ...640..691V}.  The theory of cluster magnetic fields suggests that they scale with density as $B \propto n_e^{1/2}$ \citep{2011MNRAS.410.2446K}; this is backed up by observations \citep{2010A&A...513A..30B} and is assumed by many authors \citep[e.g.][]{2015ApJ...800...60M,2015Natur.523...59M}.

\subsection{Simulation setup}\label{setup}

The numerical model used in this study is described in detail in \cite{2023MNRAS.526.3421S} (henceforth referred to as \citetalias{2023MNRAS.526.3421S}); here we sumarise the salient points.  This model was a development of that described in \cite{2013MNRAS.430..174H}, \cite{2014MNRAS.443.1482H}, \cite{2016MNRAS.461.2025E} and \cite{2019MNRAS.490.5807E} (henceforth referred to as \citetalias{2013MNRAS.430..174H,2014MNRAS.443.1482H,2016MNRAS.461.2025E} and \citetalias{2019MNRAS.490.5807E}).  The model uses the UPP atmosphere and a jet with a Lorentz factor ($\gamma = 10$) which matches well with the values seen on parsec scales.  The model also employs stretched grids in order to model the central regions at a high resolution and so enables a narrow, and so more realistic, injection cylinder.

The UPP is a self-similar profile with one input variable: cluster mass\footnote{We assume that the model cluster is at a distance of $z=0$.}; it represents observed cluster profiles more faithfully than any other model atmosphere.  The cluster mass is implemented as $M_{500}$ in units of $\text{h}^{-1}_{70}\times \num{e14}M_{\odot}$.  Following the methodology of \cite{2011MNRAS.418.1621H}, \citetalias{2014MNRAS.443.1482H}, \citetalias{2016MNRAS.461.2025E} and \citetalias{2019MNRAS.490.5807E}; we implement a cluster magnetic field which is multi-scaled, tangled and has a magnitude related to the cluster density profile.  Using the equations of the Universal Pressure Profile (see previous section), we derived the following expression for the dark matter potential, used to hold the mass of the cluster in place:
\begin{equation}
\Phi(r)=\frac{kT}{\mu m_u}\ln\left[x^{\gamma}\left(1+\left(c_{500}x\right)^{\alpha}\right)^{\left(\beta-\gamma\right)}\right],
\end{equation}
which can be compared with dark matter potentials for the $\beta$-profile \citep{2005A&A...431...45K} or the NFW-profile \citep{2008gady.book.....B}.  The simulation unit for density, length and pressure are set to $\rho_0=\num{3.01e-23}\,$kgm$^{-3}$; $l_0=2.1$\,kpc and $p_0=\rho_0c^2 = \num{2.7e-6}$\,Pa, respectively; and the simulation unit for the magnetic field is calculated from $B_0=c\sqrt{4\pi\rho_0}$, giving \num{1.84}$\,\mu$T.

The simulations are carried out on a static three-dimensional Cartesian grid centred on the origin and extending to a length of $300$ kpc in each direction.  To ensure that there are enough cells at the end of the injection cylinder to enable the jet to successfully get onto the grid, and for efficient use of computer resources, we used stretched grids.  The central patch is a 4.2 kpc cube in width and is represented by 50 grid points in the y and z-directions and 10 grid points in the x-direction.  Either side of the central patch is a geometrically stretched grid of 200 cells in the y and z-directions and 300 cells along the x-direction.  The resolution along the y and z-directions ranges from 0.084 kpc at the centre to 6.9 kpc at the grid boundary; along the x-direction the resolution ranges from 0.42 kpc at the centre to 2.1 kpc at the grid boundary.  The cell count is, therefore, $(n_x,n_y,n_z)=(610,450,450)$; all outer boundaries are set to `periodic'.  An injection cylinder is positioned in the centre of the grid, with the two jets running along both directions of the x-axis.  We use a jet radius of 0.2 simulation units (0.42 kpc).  The central patch has an aspect ratio of 5, increasing along the x-axis towards the edge of the computational domain.  The end of the injection cylinder projects beyond the end of the central uniform patch and so the cells here have an aspect ratio slightly higher than 5.  However, in our present study we are interested in the magnetic material along our line of sight (we look along the z-axis) and towards the lobes (i.e. at a distance from the centre) and these cells do not have such a large aspect ratio in comparison with those along the x-axis.  The disadvantage of using stretched grids is that greater numerical dissipation is likely to take place along the longer sides of such cells; this introduces the possibility of grid-dependant artefacts.  A limited series of tests were carried out whereby the number of cells along the x-axis was doubled (high resolution and smaller aspect ratio) or halved (low resolution and greater aspect ratio); the results can be seen in Figs 13 and 14 of \citetalias{2023MNRAS.526.3421S}.  The conclusion drawn from these tests was that the two higher resolution runs appear very similar in terms of dynamics and energy, suggesting that differences in resolution and grid anisotropy do not dominate these models.  In this paper we extended our investigation of resolution (and so cell anisotropy) to the calculation of fractional polarization and depolarization.  We find that for our two higher resolution runs, results for fractional polarization and depolarization are very similar to one another; whereas the lower resolution run achieves the same basic trends but values are significantly different to the other two runs.  This finding highlights that resolution and grid anisotropy do have an impact on the polarization results, although we believe that by using results only from the two higher resolutions, we can minimise this impact.  Comparative data for our resolution study is presented below in Section \ref{results}.

In the construction of the magnetic field of our model atmosphere, we used a power spectrum and cut the scale off below $\sim 3$ pixels (see \citetalias{2023MNRAS.526.3421S} for further details), this was motivated by RM observations that exhibit large-scale structures extending up to 100s of kpc \citep[e.g.][]{2010A&A...514A..50G, 2024MNRAS.531.2532J}.  The correlation length of the magnetic field in our models; therefore, extends over similar scales and so we are able to model RM structures and depolarization over these scales; shorter correlation lengths would not be able to model these structures and so would be less realistic.

We used {\scriptsize{PLUTO}} version 4.4-patch2 for this study \citep{2007ApJS..170..228M}; all of the runs were carried out on the University of Hertfordshire High Performance Computing facility.  Each job was run on 384 Xeon-based cores, taking between one and four weeks each. An output file was written by {\scriptsize{PLUTO}} every 50 simulation time units (every 0.34 Myr in simulation time).  We use the special relativistic magnetohydrodynamics (RMHD) physics module, HLLD approximate Riemann solvers and a second order dimensionally unsplit Runge-Kutta time-stepping algorithm, with a Courant-Freidrichs-Lewy number of 0.3.  A divergence cleaning algorithm is used to enforce $\nabla \cdot \textbf{B} = 0$.  The model assumes a single-species relativistic perfect fluid (the Synge gas) which is approximated by the Taub-Mathew Equation of State \citep{1948PhRv...74..328T,1971ApJ...165..147M}; for numerical stability reasons, shock flattening was enabled through the use of a diffusive Riemann solver (HLL) and limiter (MIN-MOD); our simulations are non-radiative for both jet and cluster material.

The jet is injected with a constant velocity of $0.994985c$; this corresponds to a Lorentz factor of $\gamma=10$, well within the range observed in jets.  We limit our study to jets with equal contributions of enthalpy and kinetic energy and inject a helical magnetic field (see Section 3.4 of \citetalias{2023MNRAS.526.3421S}).  The jet is injected with a conserved tracer quantity of value of 1.0 (and zero elsewhere).  Lobes are defined by tracer values $>10^{-3}$.  The bow shock surface (between the shock and the undisturbed ambient medium) is identified in a similar way to the tracer, by line-tracing from the edges of both sides of the volume towards the centre and finding where the radial velocity exceeds the defined value of $75$ km$~$s$^{-1}$.  In this study a suite of results was created by injecting jets of various powers into atmospheres of various masses.  Each run is described by a name of the form jetXX\_haloYY: The power of the jet had values of $0.5, 1, 2$ or $\num{4e38}$W, represented by XX having values of $05, 10, 20$ or $40$; and the mass of the atmosphere had values of $M_{500} = 0.333, 1, 3$ and $9\times\num{e14}\text{h}^{-1}_{70}M_{\odot}$, represented by YY having values of $03, 10, 30$ or $90$.  The fiducial run for this study is jet$10$\_halo$30$.

\subsection{Simulated Polarization - Methods}

The polarized emission from the lobe can be characterised by means of the Stokes parameters.  Firstly, relativistic aberration needs to be accounted for; this is where the angle between the light ray and the velocity direction in the observer's frame of reference $\theta_o$ will be different to that of the object's frame $\theta_s$ when moving at relativistic speeds.  The formula is \citep{1905AnP...322..891E}
\begin{equation}
\cos \theta_o = \frac{\cos \theta_s -\beta}{1 - \beta \cos \theta_s} 
\end{equation}
where $\beta$ is the velocity in units of the speed of light.  Therefore, we define here $B_x$ and $B_y$ as the components perpendicular to the aberration-corrected projection axis.  The Stokes $I$ (total intensity) and Stokes $Q$ and $U$ (polarized intensities) parameters are then calculated (in simulation units) by summing the following relations along the line of sight through the lobe volume
\begin{align}
&j_I=p\left(B_x^2+B_y^2\right)^{\frac{1}{2}(\alpha -1)}(B_x^2+B_y^2)D^{3+\alpha} \label{RI}\\
&j_Q=\mu p\left(B_x^2+B_y^2\right)^{\frac{1}{2}(\alpha -1)}(B_x^2-B_y^2)D^{3+\alpha} \label{RQ}\\
&j_U=\mu p\left(B_x^2+B_y^2\right)^{\frac{1}{2}(\alpha -1)}(2B_xB_y)D^{3+\alpha} \label{RU}
\end{align}
where $p$ is the local pressure, $\alpha$ is the power-law synchrotron spectral index, which is taken to be $\alpha=0.5$ corresponding to an electron energy index $p=2$ and $\mu$ is the maximum fractional polarization for a given spectral index: for $\alpha = 0.5$, $\mu=(\alpha +1)/(\alpha +5/3)=0.69$.  D is the Doppler factor, given by
\begin{equation}
D=\frac{1}{\gamma(1-\beta \cos(\theta))},
\end{equation}
where $\gamma$ is the Lorentz factor and $\theta$ is the angle between the projection vector and the velocity vector of the cell.  Here we have assumed a constant power-law spectral index which leads to the constant maximum fractional polarization quoted above, but this is an approximation and it may be that values could be as high as $100$ per cent depending upon the age of the source (see Appendix \ref{100_frac} for an explanation).

In addition to the total Stokes $I$, equation \ref{RI}, can be summed along the line of sight (ray-tracing) in order to obtain a simulated synchrotron two-dimensional emission map.  We can also create emission maps for Stokes $Q$ and $U$ (using equations \ref{RQ} and \ref{RU}). The Stokes parameters are observable quantities.  The intensity in the x and y polarization directions can be measured and from these Stokes $I$ and $Q$ calculated, similarly Stokes $U$ is found from similar measurements at $45^{\circ}$ to these axes.  The linear polarization, $\Pi$ can then be found from:
\begin{equation}
\Pi =\frac{\sqrt{Q^2+U^2}}{I}\label{linpol}
\end{equation}
and the polarization angle (otherwise known as the Electric Vector Position Angle, EVPA) can be found from
\begin{equation}
\psi=\frac{1}{2}\arctan\left(\frac{U}{Q}\right).\label{polangle}
\end{equation}
Given that the magnetic field is perpendicular to the electric field, then the observable direction of the magnetic field (projected onto the sky) $\psi_m$ is given by
\begin{equation}
\psi_m=\frac{1}{2}\arctan\left(\frac{U}{Q}\right) + \frac{\pi}{2}.
\end{equation}

\begin{figure}
\begin{center}
\includegraphics[width=0.5\textwidth]{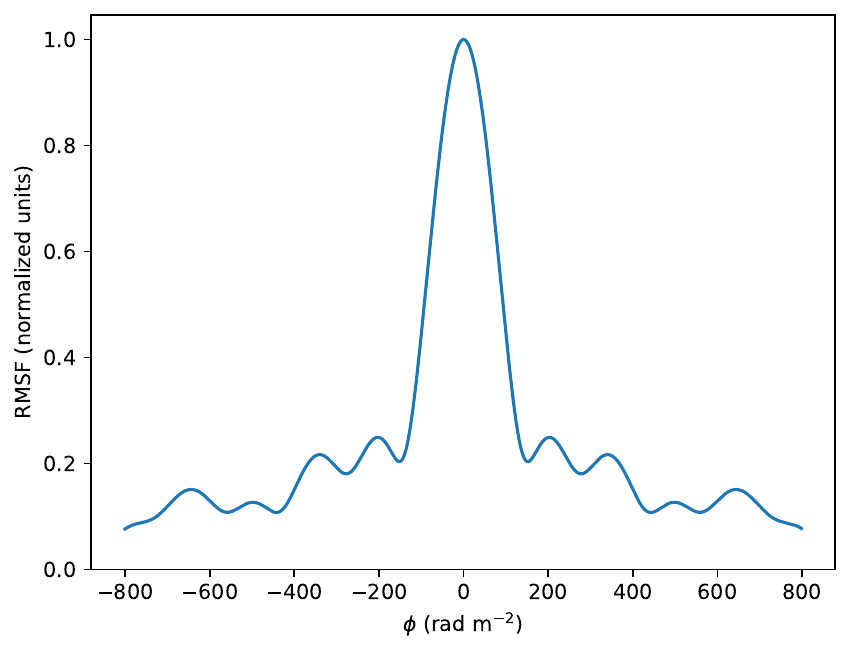}
\end{center}
\caption{Rotation measure spread function (RMSF).  Data is for the VLA with 1248 sets of $Q$ and $U$ maps - obtained by following the frequency pattern of \protect\cite{2023ApJ...955...16B}.}
\label{RMSF_Baidoo}
\end{figure}

\begin{figure}
\begin{center}
\includegraphics[width=0.5\textwidth]{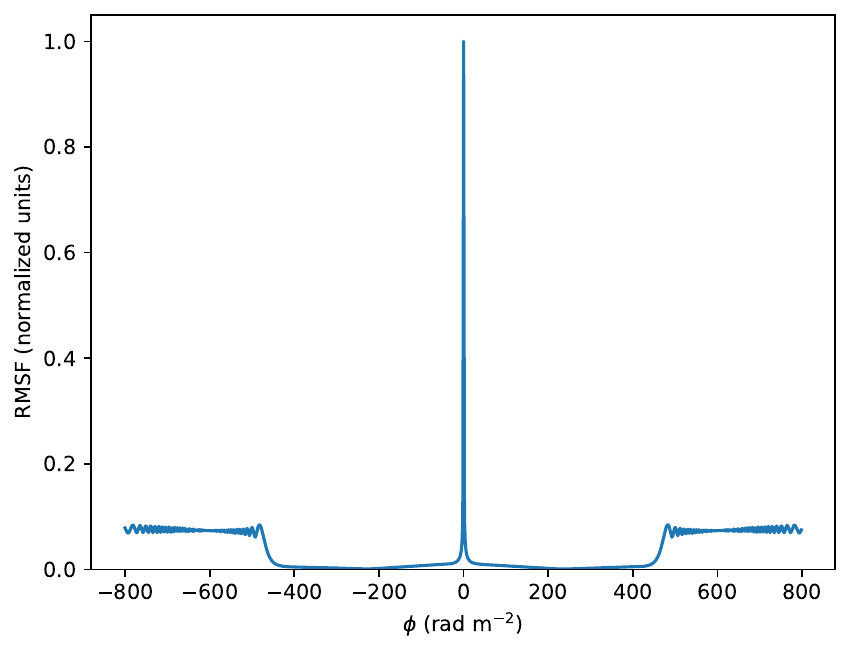}
\end{center}
\caption{Rotation measure spread function (RMSF).  Data is for 480 sets of $Q$ and $U$ maps - obtained by following the frequency pattern of LOFAR (see \protect\cite{2023MNRAS.519.5723O}).}
\label{RMSF_LOFAR}
\end{figure}

Observations of polarization are impacted by Faraday rotation.  The EVPA from the lobe emission is altered by Faraday rotation in the ICM between lobe and observer and the variability of density and magnetic field in this region leads to differential impacts upon each ray as it moves towards us and so the polarization can be reduced to very low levels, which correspond to those typically observed (as discussed above).  To obtain our depolarized fractional polarization we first generated maps of Stokes $I$, $Q$ and $U$ for the lobes (equations \ref{RI}, \ref{RQ} and \ref{RU}) as well as that of the Rotation Measure for the Faraday screen (equation \ref{RM}).  Rearranging equations \ref{linpol}, \ref{polangle} and \ref{FR} we derived expressions for the magnitudes of the simulated Stokes $Q_{\text{obs}}$ and $U_{\text{obs}}$ which would be `observed' after passing through the Faraday screen (for a particular wavelength):
\begin{equation}
Q_{\text{obs}} = \frac{\Pi\, I_{\text{int}}}{\sqrt{1+\tan^22\psi_{\text{obs}}}}
\end{equation}
\begin{equation}
U_{\text{obs}} = \frac{\Pi\, I_{\text{int}} \tan2\psi_{\text{obs}}}{\sqrt{1+\tan^22\psi_{\text{obs}}}},
\end{equation}
where $\Pi$ is the fractional polarization of the non-depolarised emission and $I_{\text{int}}$ is the Stokes $I$ emission from the lobe (the magnitude of which is not altered by depolarization).  The signs of these components can be derived from the rotated EVPA (i.e. $\psi_{\text{obs}}$), which means that our simulated measurements do not suffer from the $\pm n\pi$ ambiguity of observations.  We then use equation \ref{linpol} to obtain a map of fractional polarization after depolarization, from which mean fractional polarization (MFP) can be calculated and, for a range of Stokes $Q$ and $U$ over a range of frequencies, the RM synthesis technique can be used to obtain the Faraday dispersion functions (FDF) and so the RM (defined as the peak of the FDF).  In our study, to give concrete examples, we use the frequency patterns described by \cite{2023ApJ...955...16B} and \cite{2023MNRAS.519.5723O}; the rotation measure spread function (RMSF) corresponding to these can be seen in Figs \ref{RMSF_Baidoo} and \ref{RMSF_LOFAR}.  However, our method can be applied to any instrumental configuration and consequent RMSF.


\section{Simulated Polarization - Results}\label{results}

\subsection{Fractional Polarization Maps and Histograms}

\begin{figure*}
\begin{center}
\includegraphics[width=0.97\textwidth]{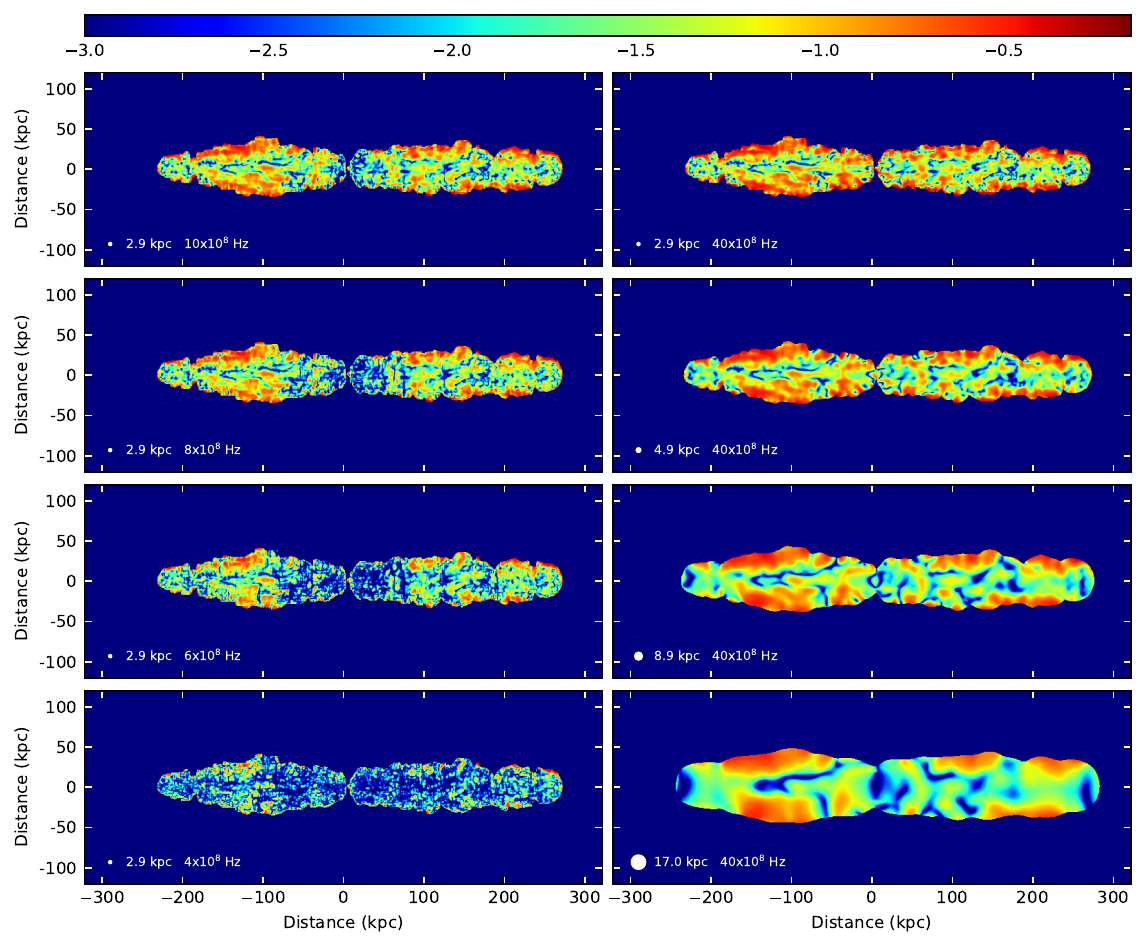}
\includegraphics[width=0.95\textwidth]{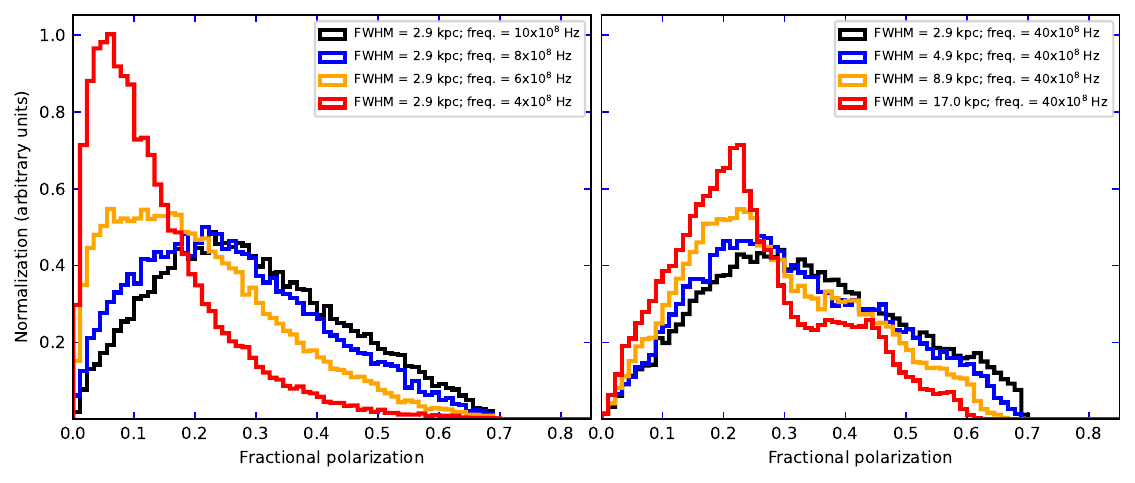}
\end{center}
\caption{Top panel: Fractional polarization images for a jet of power $\num{1e38}$W running into an atmosphere of $M_{500}=3\times\num{e14}\,\text{h}^{-1}_{70}M_{\odot}$ shown once the average lobe length has reached $250~$kpc; our high resolution run is used here (see \protect\citetalias{2023MNRAS.526.3421S}).  Left column shows decreasing observation frequency down the figure; Right column shows decreasing resolution (convolving with an increasing FWHM gaussian).  Logarithmic scale used for fractional polarization (blue is lower).  This plot is based upon Figs 3 and 7 in \protect\cite{2023ApJ...955...16B}.  Lower panel: Histograms showing the distribution of fractional polarization corresponding to the four fractional polarization images immediately above.  Histograms are labelled with the FWHM of the Gaussian used when smoothing the image and the observation frequency.  All charts for external depolarization only.}
\label{8freq_sig_images}
\end{figure*}

In order to compare our simulations with real observations we created a series of fractional polarization maps by firstly varying the frequency of observation and then by convolving with a Gaussian of increasing FWHM, see Fig. \ref{8freq_sig_images} where we have used the fiducial run of a jet of power $\num{1e38}$W running into an atmosphere of $M_{500}=3\times\num{e14}\,\text{h}^{-1}_{70}M_{\odot}$ shown once the average lobe length has reached $250~$kpc.  The fractional polarization is reduced with decreasing frequency and with reduced resolution (i.e. greater FWHM of convolving Gaussian). These trends are replicated in the histograms of Fig. \ref{8freq_sig_images} whereby it can be seen that the median polarization decreases with reducing frequency and reduced resolution.  These results are in line with expectations as depolarization is greater for lower frequencies as the electromagnetic wave is rotated more for longer wavelengths; and a lower resolution will both result in greater beam depolarization with more rays of light being combined and oppositely polarized rays combining and cancelling out.  Exactly the same pattern of results is seen for frequency and resolution in Figs 3 and 7 respectively in \cite{2023ApJ...955...16B}, where observations of the polarization of Hydra A using the Jansky Very Large Array.

\begin{figure*}
\begin{center}
\includegraphics[width=0.95\textwidth]{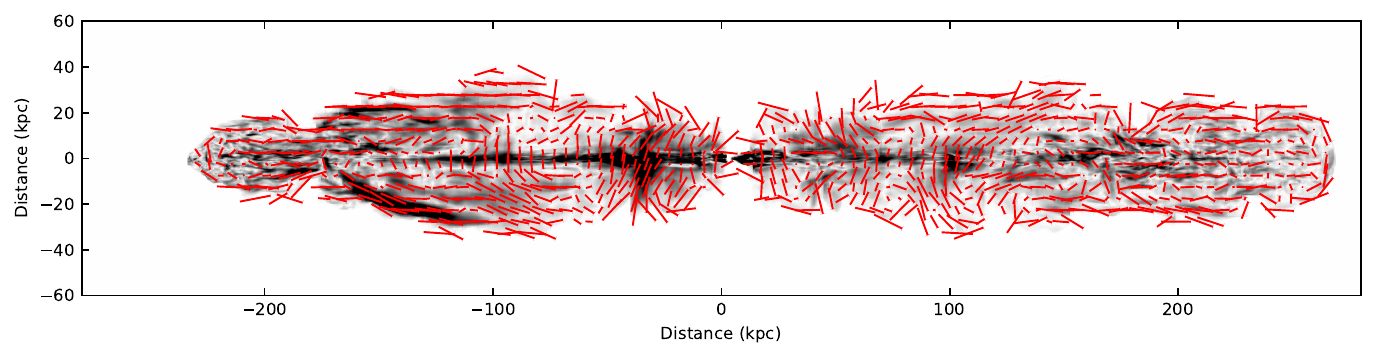}
\end{center}
\caption{A simulated radio image from the same simulation as that used in Fig. \ref{8freq_sig_images}:  The shading represents Stokes $I$ synchrotron emission and is overlaid with vectors representing the direction of the magnetic field and the magnitude of fractional polarisation.  The magnetic field direction can be seen to be (generally) parallel to the lobe-boundary and the fractional polarization tends to be higher at the lobe-boundary than in the centre of the lobes.}
\label{Highres_onechart}
\end{figure*}

A notable feature of the fractional polarization images of Fig. \ref{8freq_sig_images} is the edge enhancement, this is particularly visible in the images on the RHS.  Such enhancement at the edges of the lobe were also seen in the simulations of \cite{2011MNRAS.418.1621H} and in our previous work in \citetalias{2014MNRAS.443.1482H} (see Fig. 9).  As the lobe inflates, its magnetic field orients itself with the lobe boundary such that it becomes parallel with it, and this leads to increased polarization.  The magnetic field parallel to the lobe-boundary and the increased polarization can be seen in Fig. \ref{Highres_onechart} where the vectors' direction indicate the direction of the magnetic field and the length of the vectors indicate the relative magnitude of fractional polarization.  The vectors are predominantly parallel to the lobe boundary and their lengths at the edge are higher than in the body of the lobe.  A particular region of edge enhancement is at the top and bottom of the widest part of the left-hand lobe.  In contrast, the first $\sim 25\%$ of the right-hand lobe nearest the cluster core has a more irregular magnetic field direction on the lobe-boundary and correspondingly lower fractional polarization in the plots on the RHS of Fig. \ref{8freq_sig_images}.

\begin{figure}
\begin{center}
\includegraphics[width=0.4\textwidth]{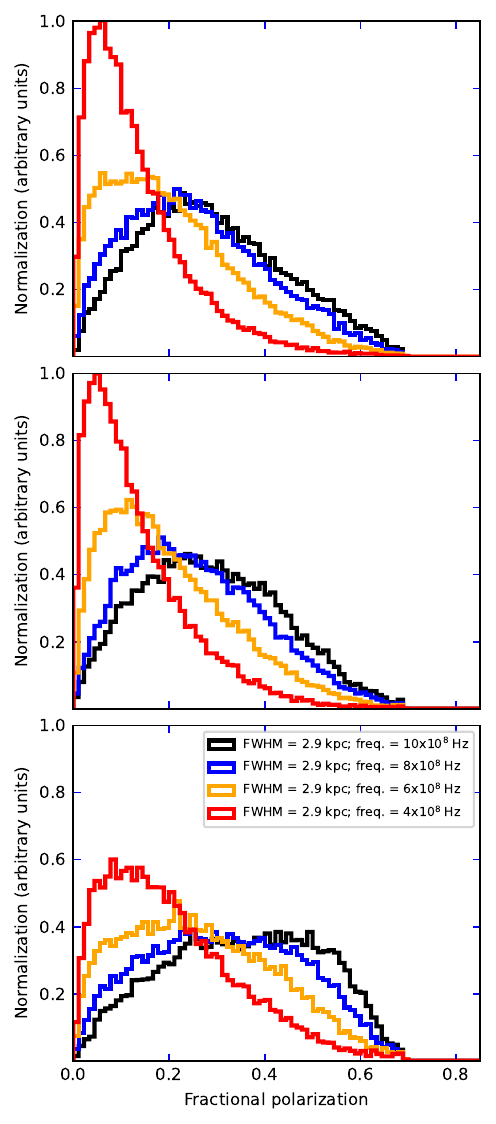}
\end{center}
\caption{Same as bottom left panel of Fig. \ref{8freq_sig_images} except for different resolution runs.  The top panel is the high resolution run (same as in Fig. \ref{8freq_sig_images}); middle panel is the normal resolution run and the bottom panel is the low resolution run.  The two higher resolution runs produce very similar results whereas the low resolution is significantly different.  The resolution of the run has an impact on the fractional polarization of the lobes, although this impact is minimised when higher resolutions are used.  Note that the low resolution run is only used as a comparator in order to establish the impact of resolution.}
\label{3x8histograms}
\end{figure}

The histograms of Fig. \ref{8freq_sig_images} are very similar to those we produced for the lower resolution runs of \citetalias{2014MNRAS.443.1482H} (which were similar to those produced by \cite{2011MNRAS.418.1621H}).  However, in \citetalias{2014MNRAS.443.1482H} we noted the presence of a considerable spike at the highest polarization in all our simulation runs, which is never present in observations.  In our current results the spike is completely removed, more than likely as a result of the more realistic, smaller injection cylinder (ten times smaller radius) and the more realistic helical injection of magnetic field (\citetalias{2014MNRAS.443.1482H} used a purely toroidal field).  This improved aspect of our model performance gives us confidence that our results are an improvement upon our previous work and, therefore, closer to realism.  As a test of how great the influence of model resolution is on our results, we replicated the bottom left panel of Fig. \ref{8freq_sig_images} for three different resolutions; the results can be seen in Fig. \ref{3x8histograms} where our two higher resolution runs produce consistent results despite differences in their resolution.

\subsection{Depolarization Maps - Frequency and Resolution}
\begin{figure*}
\begin{center}
\includegraphics[width=1.0\textwidth]{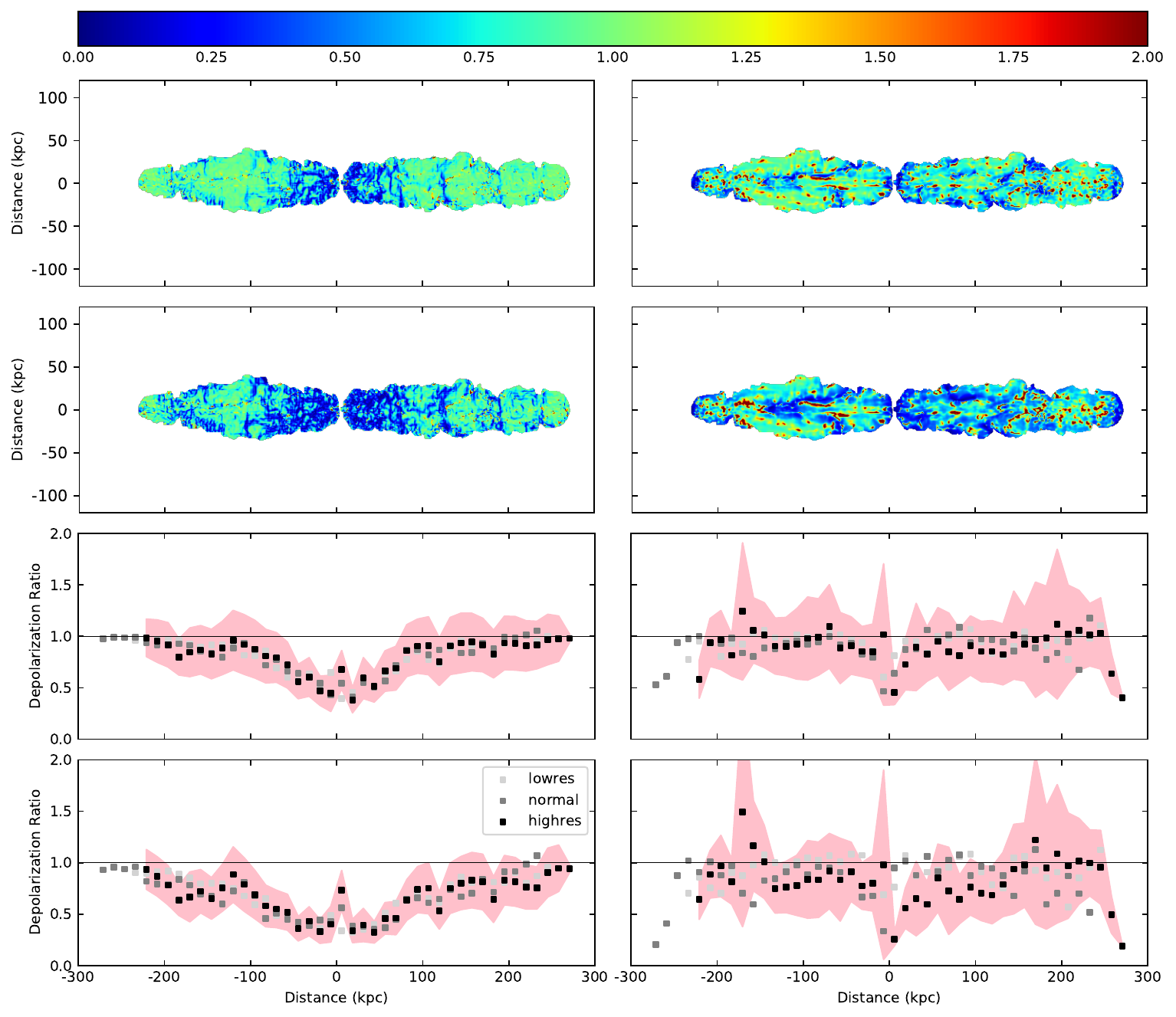}
\end{center}
\caption{Top two rows:  Depolarization ratio maps.  Left column for frequency depolarization ratio (FDR) and right column for resolution depolarization ratio (RDR).  Top FDR is the ratio of the 0.8 GHz image to the 4 GHz image; the bottom FDR is the 0.6 GHz image to the 4 GHz image.  The top RDR is the ratio of FWHM 11.6 kpc to 2.9 kpc and the bottom RDR is FWHM 23.2 kpc to the 2.9 kpc.  Bottom two rows:  Depolarization ratio charts, correspond to the top two rows.  Black squares are the average ratio, the thin black line is for ratio=1.  Pink shaded region indicates the extent of the standard deviation about the average value.  These charts are based upon Figs 4, 5, 8 and 9 in \protect\cite{2023ApJ...955...16B}.  These maps are for external depolarization only and are for a jet of power $\num{1e38}$W running into an atmosphere of $M_{500}=3\times\num{e14}\,\text{h}^{-1}_{70}M_{\odot}$ shown once the average lobe length has reached $250~$kpc; high resolution run used here (see \protect\citetalias{2023MNRAS.526.3421S}).  In the lower four panels average values for the medium (gray) and low (light gray) resolution runs are included for comparison; these demonstrate that while resolution influences depolarization, the general trends of all runs is the same.}
\label{8depol_ratio}
\end{figure*}

In order to establish the degree of depolarization, we calculate the \textit{Frequency-dependent Depolarization Ratio} (FDR) and the \textit{Resolution-dependent Depolarization Ratio} (RDR).  The FDR is found from dividing a fractional polarization map at a low frequency by one at a high frequency, whereas the RDR is found by dividing a low resolution fractional polarization map by one at higher resolution.  Examples are presented in Fig. \ref{8depol_ratio} where the top left maps are for FDR and the top right for RDR (see the caption for actual values used to create these maps).  In addition, the radial profile of depolarization is also presented in the lower half of Fig. \ref{8depol_ratio}, these correspond to the depolarization maps in the upper half of the figure.  The mean depolarization ratio is binned over a small distance and the shading represents the standard deviation.  For all these charts, lower values mean stronger depolarization; although the depolarization ratio is prone to large errors as it is a ratio of ratios and so the scatter is large, resulting in some values greater than one in our charts.  These charts present very similar results to those presented in Figs 4, 5, 8 and 9 of \cite{2023ApJ...955...16B} for observations of Hydra A, which again gives us confidence in our model.  It can be seen in Fig. \ref{8depol_ratio} that the depolarization is greatest at the centre; the explanation for this is that we have placed the AGN at the very centre of the cluster and the result is that the depolarization is greatest there as it is a richer environment.  We can, therefore, use our depolarization maps as a proxy for the density distribution of our cluster environment (given we are observing our jets at an angle of $90^{\circ}$ to the jet axis).

\subsection{The Laing-Garrington Effect}\label{lg}
Depolarization impacts the lobe created by the jet which is further away from the observer (the counterjet) more than the lobe created by the jet moving towards us (the jet), this was established by \cite{1988Natur.331..149L} and \cite{1988Natur.331..147G} and is known as the Laing-Garrington effect.  The simple explanation is that there is a greater depth of Faraday screen between the observer and the more distant counterjet (all other factors in equation \ref{RM} being equal), and so the radiation from that more distant lobe will experience a greater degree of depolarization in its journey to the observer, and so have a lower polarization.  Using the Laing-Garrington effect, \cite{1993ApJ...416..554T} estimated the inclination of Hydra A to be $\leq 60^{\circ}$.

\begin{figure*}
\begin{center}
\includegraphics[width=1.0\textwidth]{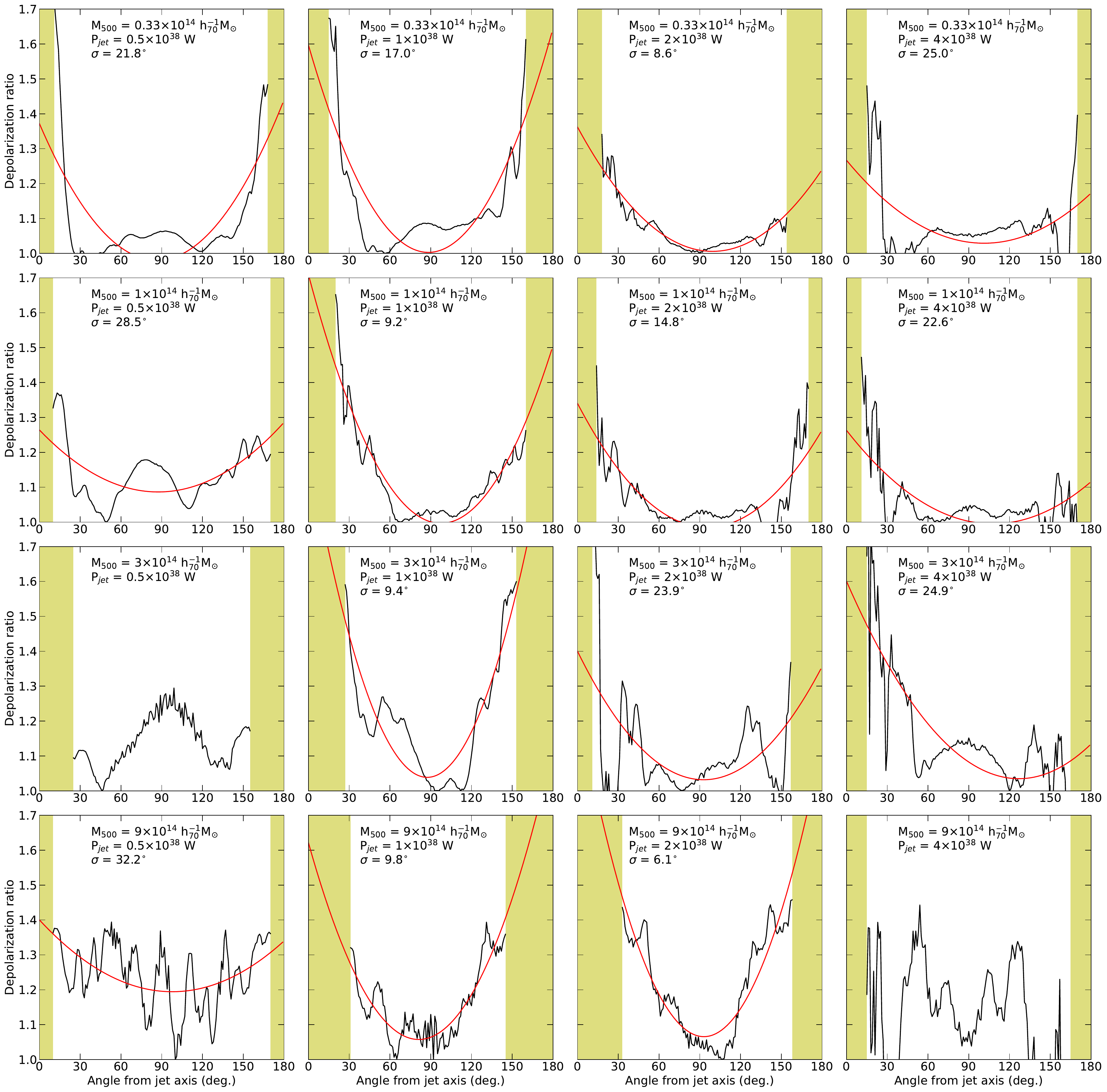}
\end{center}
\caption{Variation of the Laing-Garrington effect with angle of view for the full range of cluster mass and jet power used in this study.  An angle of $0^{\circ}$ corresponds to looking along the jet axis, $90^{\circ}$ is when the AGN is in the plane of the sky and $180^{\circ}$ is when we look along the counterjet.  The density of the cluster atmosphere increases down the chart from top to bottom as $M_{500} = 0.333, 1, 3$ and $9\times\num{e14}\text{h}^{-1}_{70}M_{\odot}$, and the jet power increases from left to right as $0.5, 1, 2$ and $\num{4e38}$W.  The observation frequencies and resolution of our model match those used by \protect\cite{1991MNRAS.250..171G}.  The black line is the depolarization ratio (i.e. Laing-Garrington effect) and the red line is a quadratic fitted to the data using a least-squares technique.  The standard deviation between these two lines has been calculated for each graph (and labelled as $\sigma$).  The areas shaded in yellow are where the overlap between the the two lobes means that the Laing-Garrington effect can no longer be calculated accurately, these areas are excluded from the calculation.}
\label{16LG_chart}
\end{figure*}

In a similar way to our previous work in \citetalias{2014MNRAS.443.1482H}, we simulate observations based on those described by \cite{1991MNRAS.250..171G}:  We derive the `observed' Stokes maps after depolarization (i.e. $Q_{\text{obs}}$ and $U_{\text{obs}}$) at frequencies of $1.4$ and $5$ GHz for each lobe separately.  Before generating our resolved fractional polarization maps we convolve our simulated Stokes parameter maps with a Gaussian designed to represent roughly 15 beam widths across the source (thus matching the observations of \cite{1991MNRAS.250..171G}).  The depolarization is calculated as the ratio of mean fractional polarization at each frequency (DP = $f_{1.4}/f_{5}$) and is found for each lobe separately.  We then find the depolarization ratio, DPR = DP$_{\text{j}}/$DP$_{\text{cj}}$ , where the subscript indicates either `jet’ or `counterjet’.  The depolarization ratio defined in this way is expected to be larger than unity, although the multi-scaled and tangled nature of the model ambient magnetic field will ensure significant variability about any trend for this value (as seen in \citetalias{2014MNRAS.443.1482H}).  We replicate these results for our current model by plotting the variation of the depolarization ratio with viewing angle for the full range of our cluster masses and jet powers (see Fig. \ref{16LG_chart}).  The LG effect will be impacted by the multi-scaled and tangled nature of our magnetic field and the angle of view will determine which parts of the spatially-varying field the LOS passes through and so we expect considerable scatter in our charts (as was noted in our previous work, \citetalias{2014MNRAS.443.1482H}, Fig. 12).

In Fig. \ref{16LG_chart} we have plotted the LG effect as the black lines and, as expected, there is considerable variability and the depolarization ratio tends to have a minimum around $90^{\circ}$ and increases away from this towards the jet-counterjet axis.  We do not know the underlying function which determines the relation between the Laing-Garrington effect and the angle of view; but in an attempt to quantify the scatter in our data, we have made the assumption that the shape of this curve is a quadratic (this has been determined simply from the observations presented here).  Using a least-squares technique, we fit a quadratic curve to these results (shown in red) and the standard deviation is calculated between the depolarization ratio and this best-fit quadratic curve.  It can be seen that two of our charts are so poorly behaved (as a result of the highly variable magnetic field) that a quadratic curve would not fit.  The other 14 curves have uncertainties in the range $\sigma \sim 6^{\circ}$ to 33$^{\circ}$.  The general trend appears to be that the greater the cluster mass atmosphere, the greater uncertainty in the angle, which can be explained in terms of a greater (and consequently more variable) magnetic field (given the magnetic field scales as $\sqrt{n_e}$ in our models, see Section \ref{setup}).  We conclude that, within our simulations, we can use the Laing-Garrington effect in order to constrain the angle of view to $\sim 10^{\circ}$ (at best).  It would be interesting to see whether such constraints could be derived observationally for a large sample of radio sources.

\subsection{Mean Fractional Polarization - Variation with Frequency and Resolution}
\begin{figure*}
\begin{center}
\includegraphics[width=0.92\textwidth]{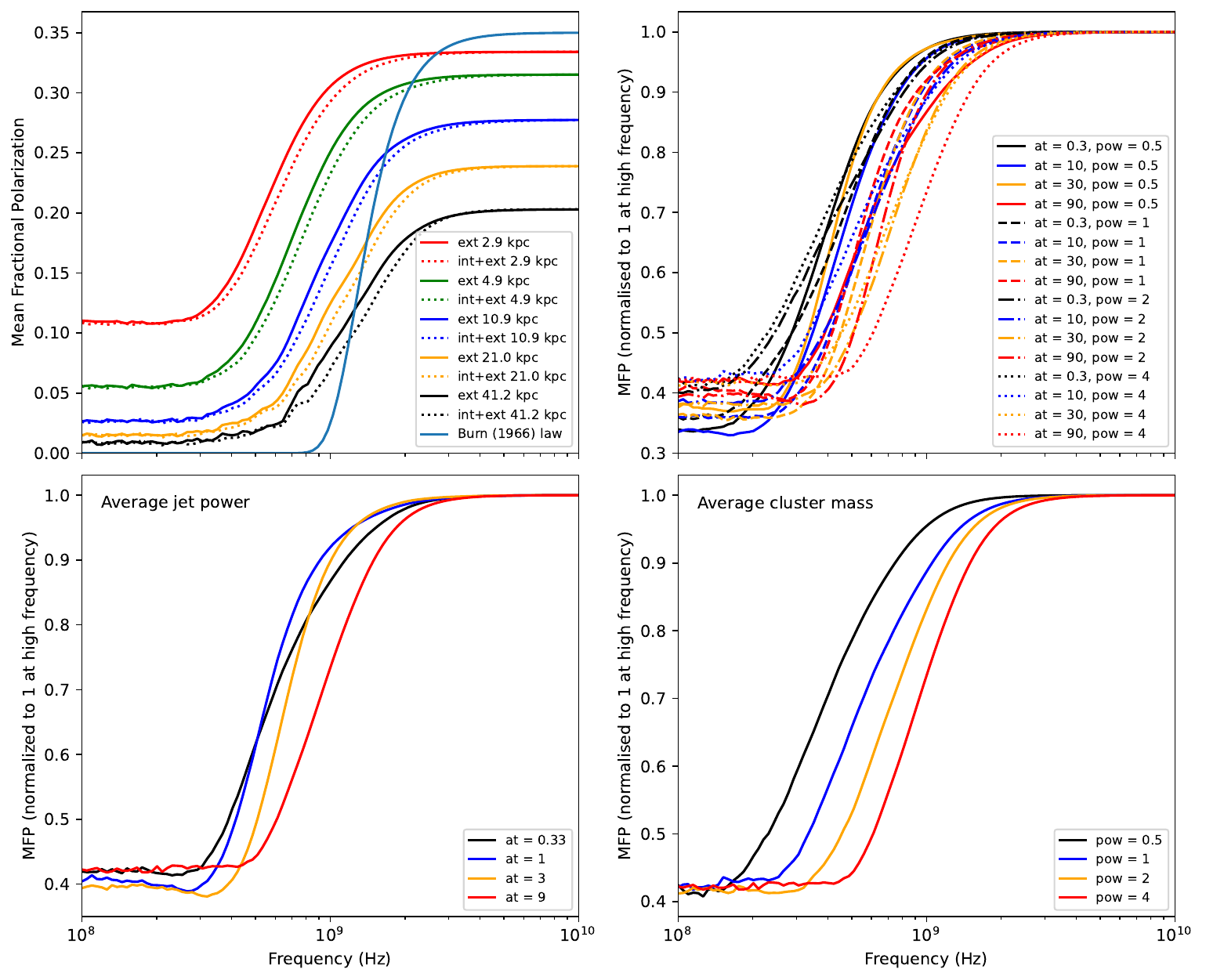}
\includegraphics[width=0.92\textwidth]{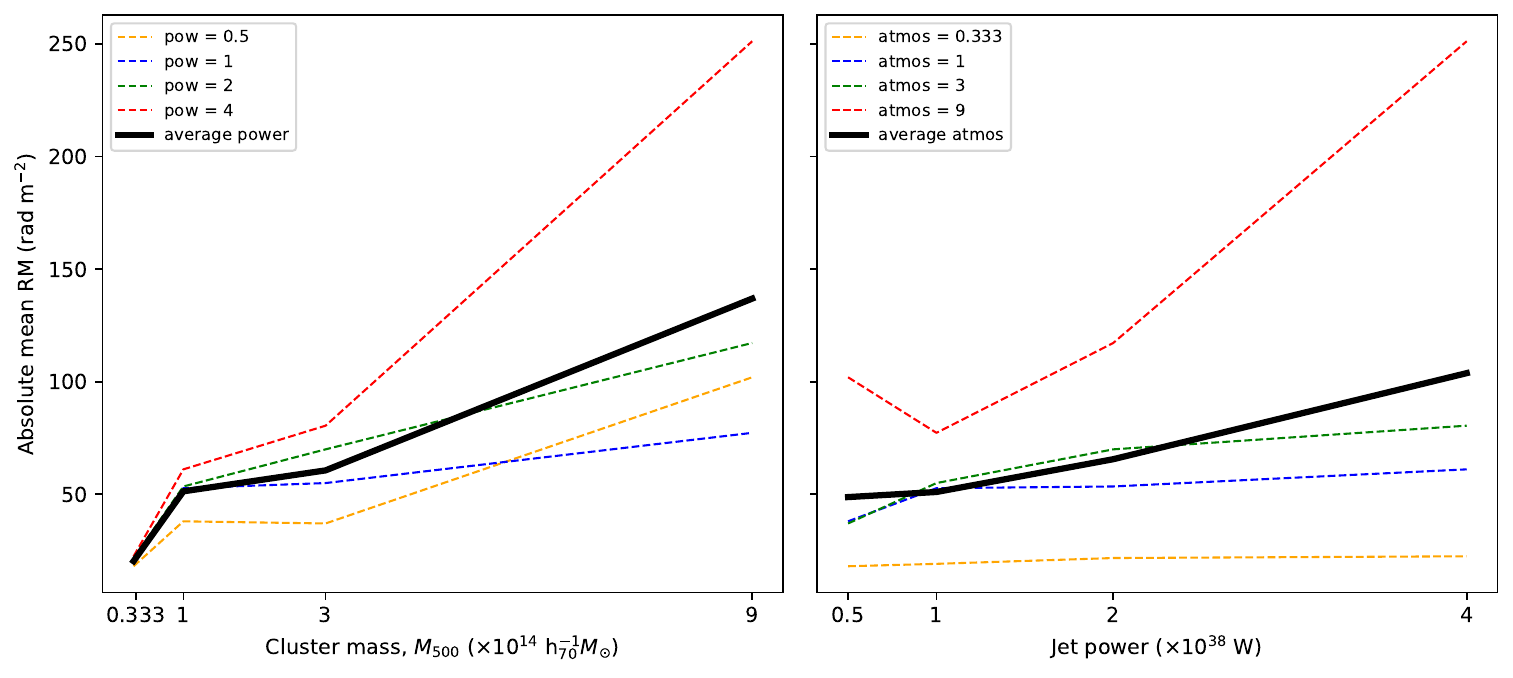}
\end{center}
\caption{Top left:  Mean fractional polarization (MFP) as a function of frequency for a range of resolutions (as defined by the FWHM of the Gaussian used to convolve with the original image) for a jet of power $\num{1e38}$W running into an atmosphere of $M_{500}=3\times\num{e14}\,\text{h}^{-1}_{70}M_{\odot}$, shown for external depolarization only and for both external and internal polarization; also shown is the Burn (1966) law (see text).  Top right:  MFP for all runs (external polarization) normalised to the value of the highest frequency; cluster mass is given as $M_{500}$ and measured in units of $\times\num{e14}\,\text{h}^{-1}_{70}M_{\odot}$ and jet power is in units of $\times\num{e38}$W.  Middle row:  Same data as top right panel except that runs have been averaged for a fixed jet power (left panel) and fixed cluster mass (right panel), this demonstrates that higher density and lower power runs have lower MFP - corresponding to rounder lobes, suggesting that lobe morphology influences MFP.  Bottom row:  Absolute mean RM for all runs by power and atmosphere; these suggest that increases in cluster mass and jet power both increase the RM (see text).  All values are for once the lobes have reached an average length of 250 kpc.  With the exception of the top left chart, all runs are for a resolution of FWHM $2.9$ kpc.}
\label{mfp}
\end{figure*}

The Mean Fractional Polarization (MFP) is found by finding the average non-zero value for a fractional polarization map.  Given that polarization varies with frequency and resolution, we expect the MFP to also vary with frequency and resolution, as can be seen from our top left panel of Fig. \ref{mfp} where MFP is reduced for lower frequencies and lower resolutions.  A decrease in resolution reduces the MFP; this is to be expected as a lower resolution results in a greater amount of beam depolarization as regions of positive and negative polarization within the beam will be averaged out.  Depolarization is considered to be due to an unresolved or partially resolved foreground screen, and can largely be avoided by using high resolutions and high frequencies \citep[e.g.][]{2012MNRAS.423.1335G}; although the existence of small scale structure and limited resolution (particularly of distant objects) means that it is not always possible to avoid some depolarization, as is the case for the depolarized galaxies discussed by \cite{2010MNRAS.401.2697H, 2012MNRAS.424.1774H}.

In order to model the external polarization, we employ the Burn Law (equation \ref{Burn_law}):  In the top left panel of Fig. \ref{mfp} we have plotted the curve predicted by assuming an arbitrary value of $d$ to be the size of one pixel, $p_i = 0.35$ and other values are taken from our numerical model as average values for the shocked region.  The resultant Burn law gives a shape remarkably similar to those found from our results, although it predicts a faster fall towards zero as the wavelength decreases.  In fact, our results show that the mean fractional polarization does not reach zero, particularly at high resolution, presumably because there are some lines of sight that have very little effective depolarization. A general prediction of this type of model is that the fractional polarization will not show the exponential cutoff to zero of the Burn law, as in fact is observed in some cases \citep[e.g.][]{2003MNRAS.339..360H}. Moreover, even though the Burn law describes the qualitative shape of the depolarization curve well, equation \ref{Burn_law} does not take account of the beam size and magnetic field power spectrum, both of which will impact how quickly the value falls to minimum as the wavelength decreases.  The impact of beam size can be seen in our results but we have used only one value for the power spectrum in our models and so further work would need to be done to demonstrate its impact.

In addition to the external polarization, we also produced results for combined external and internal polarization (dotted lines on top left panel of Fig. \ref{mfp}) where internal depolarization includes depolarization caused by material inside our lobe (identified by the tracer, see Section \ref{setup}) and external is due to everything else from the nearest lobe surface. Just like the material in the Faraday screen, the internal depolarization depends upon the density and magnetic field values experienced by the emitted radiation.  Given that the density within the lobes is significantly lower than in the material external to the lobes, we find that the internal depolarization has significantly less impact than the external depolarization (for the resolution we have chosen); this is observed in the top left panel of Fig. \ref{mfp} where the MFP is shown for external only and for combined internal and external depolarization.  The smaller effect of internal depolarization is also seen in Figs \ref{morph_atmos} and \ref{morph_power} where the internal polarization is plotted separately and can be seen to have much less impact in comparison with the external. There is little evidence of internal depolarization in observations of FRII radio galaxies, it is more likely that we would find evidence of internal depolarization in FRI sources as we expect them to have entrained a greater quantity of cluster gas into their lobes, resulting in a higher lobe density.  \cite{2013ApJ...764..162O} have suggested that there is evidence of depolarization in the lobes of Centaurus A, and predict a density of lobe material of \num{e2} m$^{-3}$, which is about an order of magnitude higher than the average density of our FRII model lobes.  And so, given our current observational limitations, we are unlikely to confirm internal depolarization in the lobes of FRII's at this time.

\subsection{Mean Fractional Polarization - Influence of cluster mass and jet power}\label{MFP_cmjp}
\begin{figure*}
\begin{center}
\includegraphics[width=1.0\textwidth]{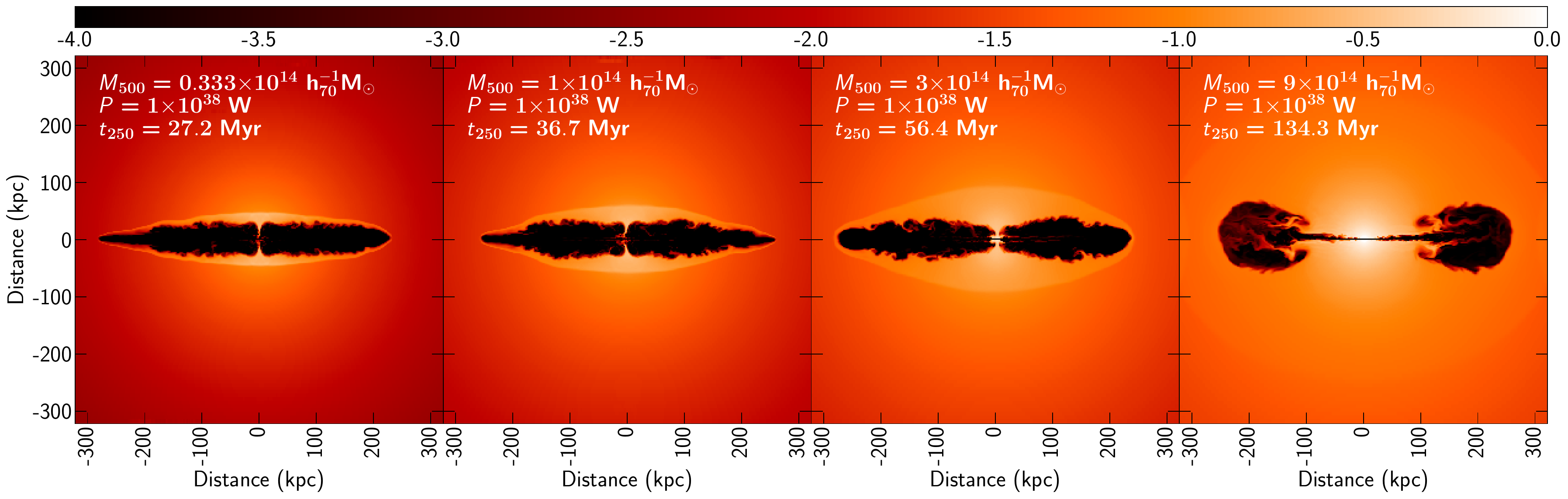}
\includegraphics[width=0.98\textwidth]{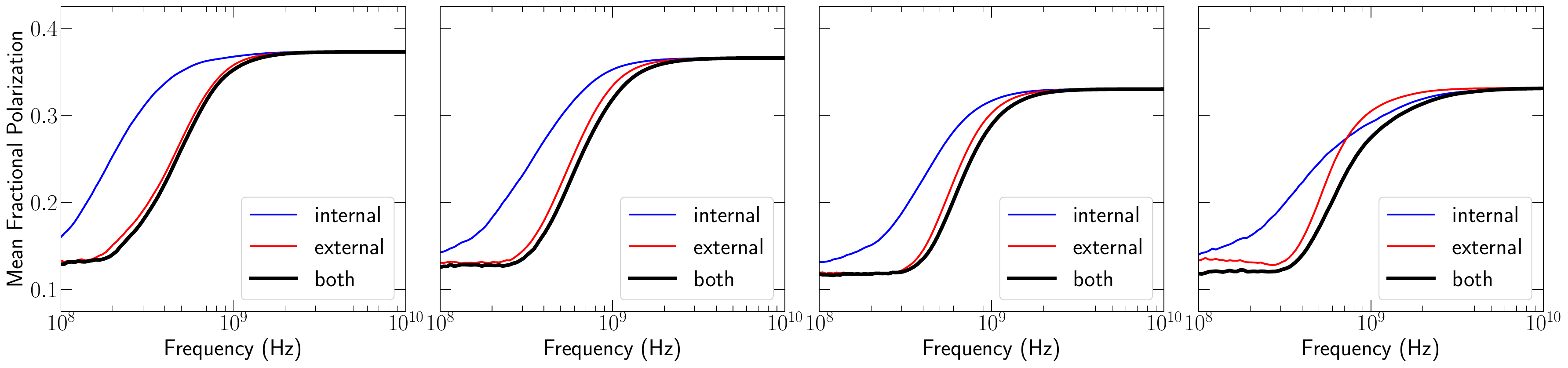}
\end{center}
\caption{Top panel: Cross-section through a logarithmic density plot of various cluster atmospheres (labelled); Lower panel:  The corresponding evolution of Mean Fractional Polarization (MFP) with frequency.  As indicated, runs for internal, external and combined polarization.  All results convolved with a Gaussian of FWHM of 2.1 kpc and for models extended to an average lobe length of 250 kpc.  Higher cluster mass results in greater depolarization (see text).}
\label{morph_atmos}
\end{figure*}

\begin{figure*}
\begin{center}
\includegraphics[width=1.0\textwidth]{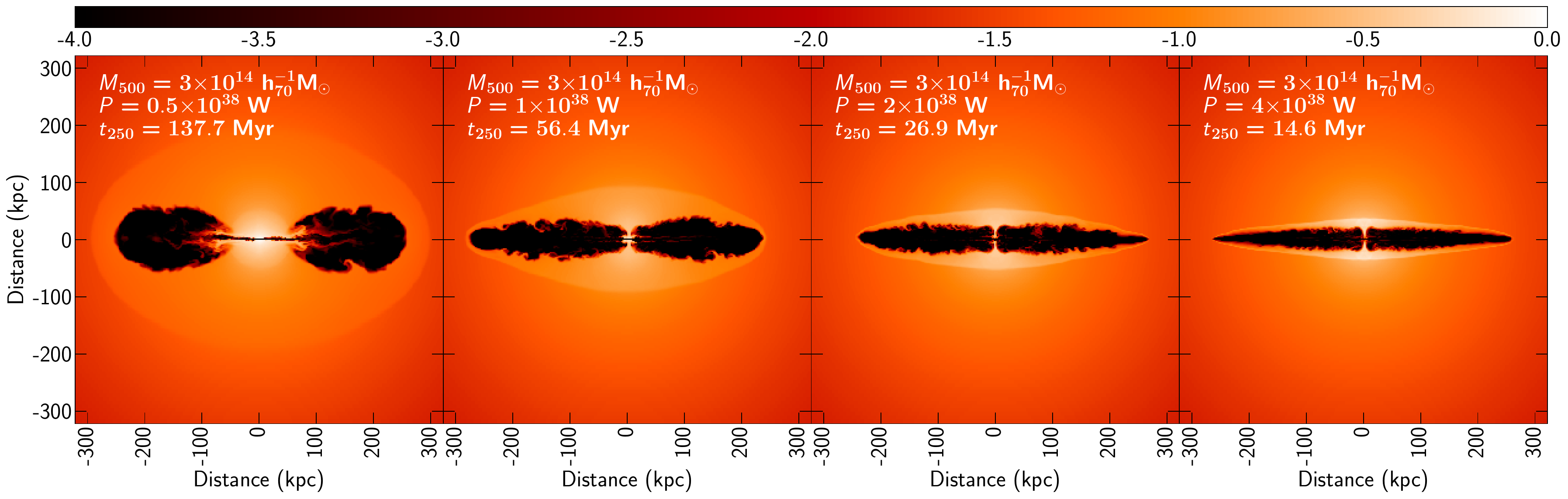}
\includegraphics[width=0.98\textwidth]{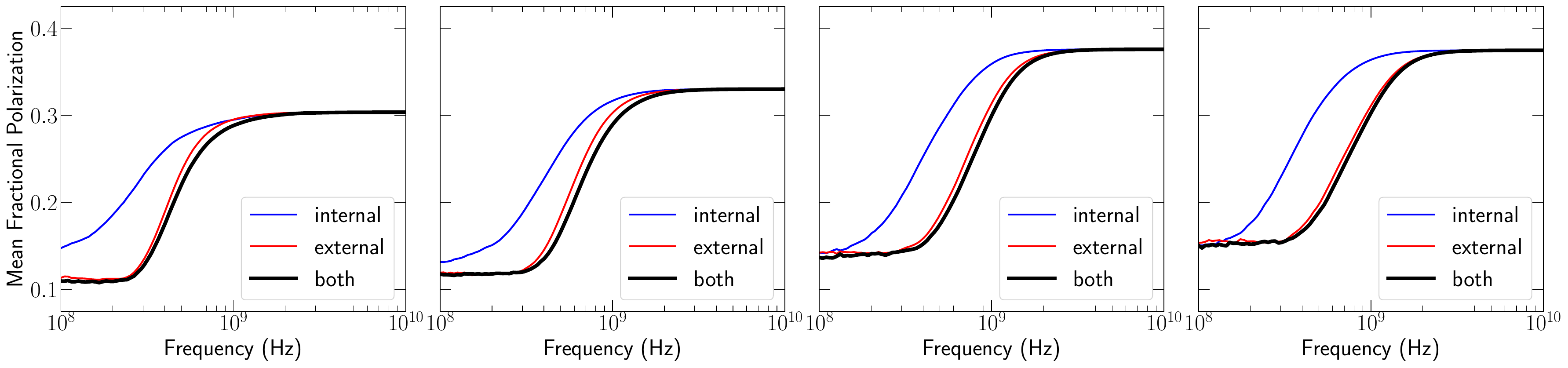}
\end{center}
\caption{Top panel: Cross-section through a logarithmic density plot of various powers (labelled); Lower panel:  The corresponding evolution of Mean Fractional Polarization (MFP) with frequency.  As indicated, runs for internal, external and combined polarization.  All results convolved with a Gaussian of FWHM of 2.1 kpc and for models extended to an average lobe length of 250 kpc.  Higher power runs result in greater depolarization (see text).}
\label{morph_power}
\end{figure*}

The top right panel of Fig. \ref{mfp} shows the evolution of MFP with frequency for the full range of cluster mass (measured as $M_{500}$) and jet power used in our suite of models; these are presented with high values normalized to unity for ease of comparison.  It is clear that cluster mass and jet power influence the fractional polarization.  In order to separate out these two factors, two further plots were created and are presented in the middle row of Fig. \ref{mfp}.  The left chart averages out the four jet powers at each cluster mass, the right chart averages the four cluster masses at each jet power.  It is clear that jet power has a highly consistent impact upon the MFP:  The greater jet power, the lower the MFP (i.e. greater depolarization) at a specific frequency.  Similarly for cluster mass, the trend is that the greater the cluster mass, the lower the MFP for a specific frequency.  The trend in MFP with power and cluster mass can be linked to the morphology of the lobes.  In Fig. \ref{morph_atmos} (top panel) it can be seen that higher mass clusters lift the lobes more from the centre through buoyancy and at the same time the jets' journey away from the core is impeded more so that the lobes become separated, inflated and well-rounded.  In Fig. \ref{morph_power} (top panel) it can be seen that higher jet power produces higher aspect-ratio lobes, the jet moves forward more rapidly and the cluster atmosphere does not have time to lift the inner lobe away from the core; this results in very pointed lobes.


In order to investigate the links found between MFP and lobe morphology, the RM of the Faraday screen was analysed.  The RM maps for each run, once the average lobe length had reached 250 kpc, were created and the mean absolute RM for each map was found; plots are presented in the bottom row of Fig. \ref{mfp}.  An increase in cluster mass corresponds to a greater mean absolute RM (left plot); this supports the findings above as greater values of RM would lead to greater depolarization.  Greater values of RM are to be expected as there will be higher values for both density and magnetic field strength in the higher mass cluster.  The variation of absolute mean RM with jet power; however, is less clear.  The conclusion is that jet power increases the mean absolute RM, although the data is not secure as only the highest mass run convincingly shows this.  Nevertheless, a trend of increasing RM with jet power can be explained in terms of the morphology of the lobes: Higher power jets progress faster and so the material near the core has less time to move outwards through buoyancy before the lobe extends to our average 250 kpc length; the result is that a greater proportion of high power jets' lobes remain near the core and so when we observe the lobes, we ray-trace through more of the inner `high RM' material and so the mean absolute RM will be higher.  
Another factor which impacts the mean absolute RM will be the impact on RM of the expanding lobes.  When lobes are inflating, they push back the surrounding atmosphere and so increase its density and magnetic field strength; the result is that the rotation measure is increased at the lobe boundaries; this was demonstrated by the work of \cite{2011MNRAS.418.1621H} where higher RM values were seen in models with inflated lobes in comparison with models without.  This effect will contribute towards the increased RM with increased cluster mass as higher mass clusters have more rounded lobes; however, this effect acts in the opposite way to our link between RM and jet power, as lower jet powers have more rounded lobes (see Fig. \ref{morph_power}).  We must conclude that the impact of the expanding lobe on increasing the RM is a smaller effect than our explanation above of higher jet powers having more of the lobe in the central regions and so increasing the mean absolute RM.


In order to model the contribution of depolarization within the lobes, we calculate the critical frequency from equation \ref{crit_freq} (the frequency below which internal depolarization becomes important).  We find that values increase monotonically with cluster mass (where jet power has been averaged) in the range $0.2-1$ GHz, so agreeing with the pattern seen in Fig. \ref{morph_atmos}, where the internal polarization is plotted separately from external and the frequency of the onset of appreciable depolarization increases with cluster mass.  Fig. \ref{morph_power} shows the sequence for increasing power, which also shows a trend of increasing depolarization with increased jet power (although this is not as easily discerned as for cluster mass).  However, the calculated critical frequency (for average cluster mass) does not give a clear trend for increasing jet power but somewhat erratic results between $0.3 - 0.7$ GHz.  In calculating the critical frequency, we must remind ourselves that in constructing our model, assumptions about the values within the lobe were made as these are poorly constrained by observations.  In addition, the formula used (equation \ref{crit_freq}) leaves out much of the physics of real jets.

\subsection{Rotation Measure Visualisation}
\begin{figure}
\includegraphics[width=0.42\textwidth]{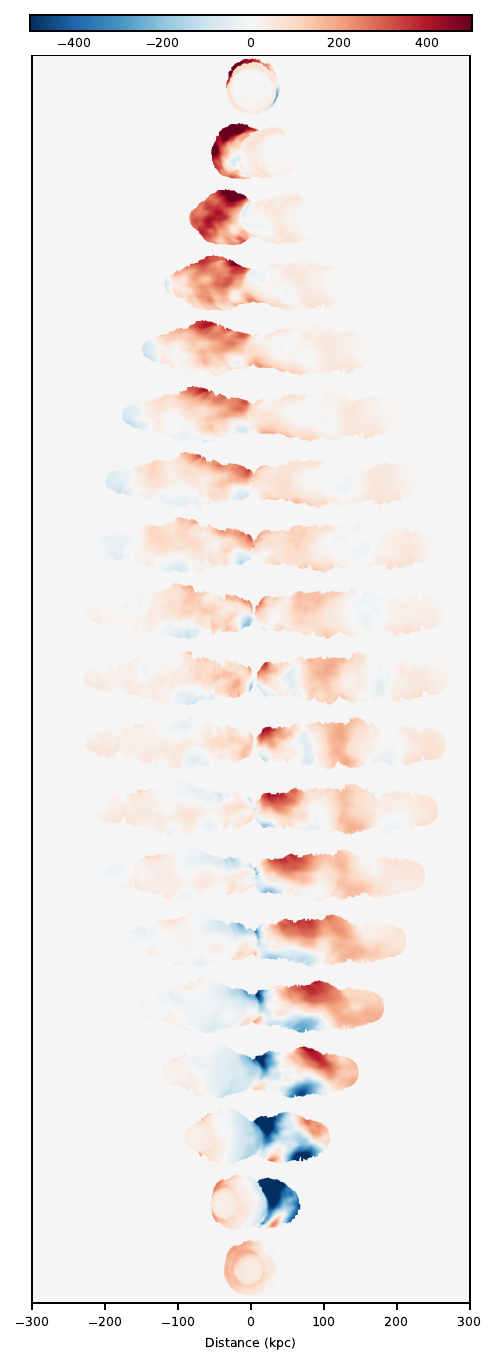}
\caption{Sequence of RM maps; top image observing down the jet axis, each subsequent image is positioned an additional $10^{\circ}$ from the axis, so that the bottom image is looking down the counterjet. Images for a jet of power $\num{1e38}$W running into an atmosphere of $M_{500}=3\times\num{e14}\,\text{h}^{-1}_{70}M_{\odot}$ with jet extended to an average length of 250 kpc.  Larger values of RM are visible when viewing the counterjet at smaller angles to the jet-counterjet axis as the more distant lobe is viewed through higher values of magnetic field, density and depth of cluster material.  Note: High resolution run used here (see \protect\citetalias{2023MNRAS.526.3421S}).}
\label{multiple_rm_images}
\end{figure}

We create Rotation Measure (RM) maps by use of equation \ref{RM}, ray-tracing along our line of sight from the lobe-shock boundary to a fixed distance of $300$ kpc from the centre of the datacube.  Using our fiducial run for a jet of power $\num{1e38}$W running into an atmosphere of $M_{500}=3\times\num{e14}\,\text{h}^{-1}_{70}M_{\odot}$, we created a sequence of RM maps at intervals of $10^{\circ}$, starting by looking down the jet and finishing by looking down the counterjet after a rotation of 180$^{\circ}$ (see Fig. \ref{multiple_rm_images}).  It can be seen that values of RM are considerably higher when viewing the lobe inflated by the counterjet at a shallow angle; this is expected as the line of sight passes through regions close to the core of the AGN which have higher values of density and magnetic field strength, as well as a greater depth of view (in comparison with viewing the nearby lobe).  This variation of RM with angle is also seen in our a fly-by movie showing the RM map around the entire 360$^{\circ}$ rotation of our higher resolution model: \url{http://uhhpc.herts.ac.uk/~ms/rm.html}.  These images (and movie) enable a dynamic visualisation of both RM and also the Laing-Garrington effect (see Section \ref{lg}).


\begin{figure}
\includegraphics[width=0.5\textwidth]{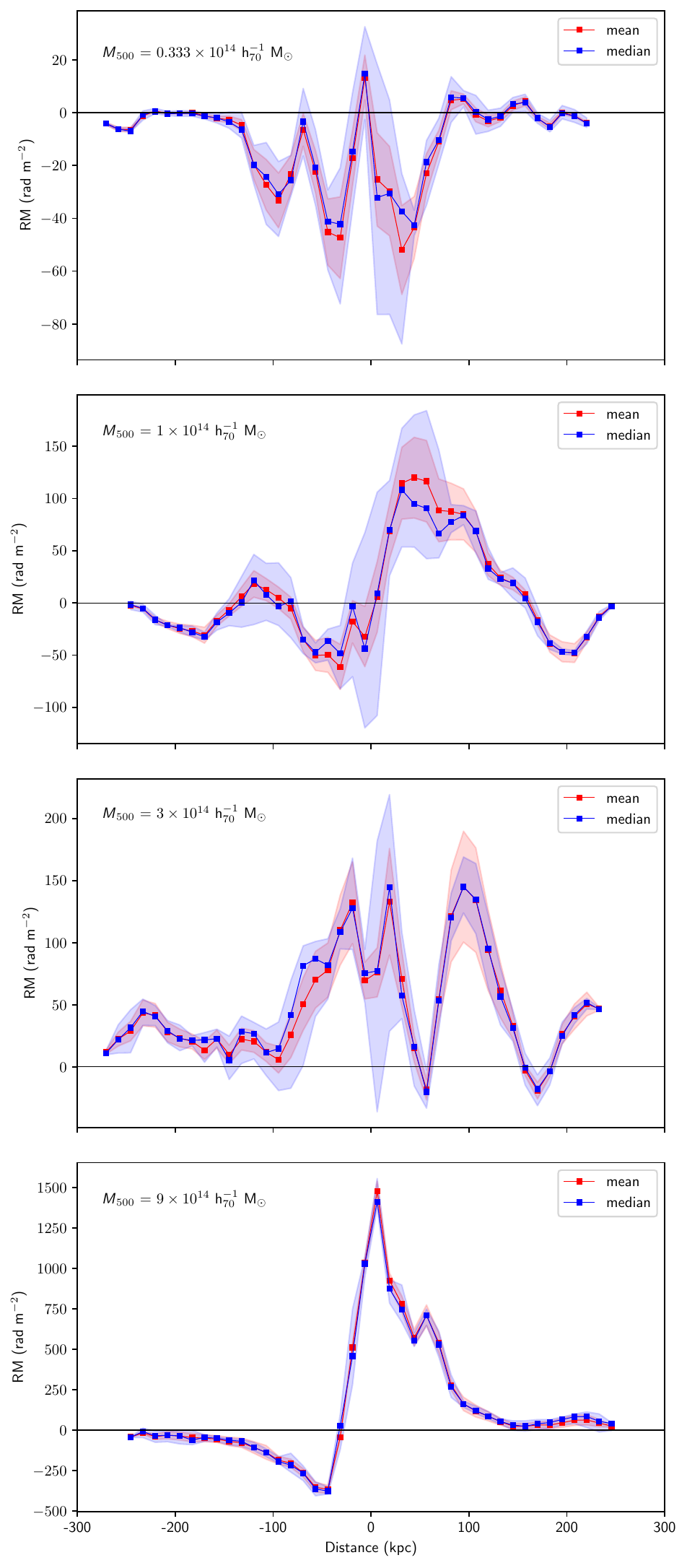}
\caption{The RM profile as a function of distance from the AGN core for different cluster atmospheres.  The cluster mass is: $M_{500}=0.333$ (top panel)$, 1, 3$ and $9\times\num{e14}\,\text{h}^{-1}_{70}M_{\odot}$ (bottom panel).  The jet has a power of $\num{1e38}$W and lobes are extended to a fixed average length of 250 kpc.  Data are gathered in bins of width 12.6 kpc.  The mean is shown in red data points, and the standard deviation is in a red shade.  The median is shown in blue data points, and the first and third quartiles are in the blue shade.}
\label{RM_distribution_with_x}
\end{figure}

Fig. \ref{RM_distribution_with_x} shows the RM profiles of the lobes along their length.  These profiles were derived from our simulated RM maps; values at a distance along the jet axis were binned in widths of 12.6 kpc (15 pixels).  In red we show the mean (data points) and standard deviation (shade), and in blue is the median (data points) and the upper and lower quartiles (shade).  The cluster mass increases for each panel going down Fig. \ref{RM_distribution_with_x} (note the different scales on the y-axis) and towards the centre of the clusters.  These cluster atmospheres were created independently of one another and so the magnetic field distributions vary considerably, the multi-scaled nature of these variations being visible (see \citetalias{2023MNRAS.526.3421S} for details).  Given that the distance from the lobes of these runs to the edge of the computational domain is roughly constant (c. 300 kpc), it is the variation of cluster density and magnetic field which leads to the variations in RM.  The jet power also has an effect, although this is less significant.  It has a role in altering the shape of the lobes and so the size of the Faraday screen, as well as a small effect in amplifying the magnetic field and increasing the density through pushing the surrounding material back upon itself, as described by \cite{2011MNRAS.418.1621H} and discussed in Section \ref{MFP_cmjp}.  These charts can be compared with those from observations, such as those of \cite{2023ApJ...955...16B}; both ours and those of \cite{2023ApJ...955...16B} demonstrate considerable variations in RM value with distance from the cluster centre, often moving between positive and negative values, indicative of the multi-scaled magnetic field believed to be a characteristic of clusters.

\subsection{RM Synthesis of 3C218, VLA Observations}\label{Baidoo_RMsynthesis}
Hydra A (3C218) is a highly-luminous Fanaroff-Riley type I (FR I) radio galaxy and has been the subject of a number of polarization studies \citep[e.g.][]{1990ApJ...360...41T, 1991MNRAS.250..171G, 1993ApJ...416..554T, 2005A&A...434...67V, 2023ApJ...955...16B}.  Here we use our model simulation as a representation of 3C218 by placing it at the same distance as 3C218 and setting the resolution and noise level to match those seen in the observations of \cite{2023ApJ...955...16B}; but we note that our cluster and radio source properties will be different to those of 3C218.

Using the values from table \ref{noise_values}, we can make use of the `scale' which converts from distance-size of object to angular-size and so convert from beam resolution given in arcseconds by \cite{2023ApJ...955...16B} to distance in kpc on the object and number of pixels in our model.  We can also compare the solid angle of the beam ($\sigma = \pi \theta_1 \theta_2 /4 \ln 2$) to the solid angle of one pixel in our model in order to find the number of pixels in the beam.  We then calculate the Stokes $Q$ and $U$ channel noise in simulation units as
\begin{equation}\label{noise}
\left(\frac{\text{pixel model noise}}{\text{sim. pix$^{-1}$}} \right) = \left(\frac{\text{source noise}}{\text{Jy pix}^{-1}} \right) \times \frac{F_{\text{M}}}{F_{\text{S}}},
\end{equation}
where $F_{\text{M}}$ and $F_{\text{S}}$ are the fluxes of the model and source in simulation units and Jy respectively and the Stokes $Q$ and $U$ channel noise has been converted from Jy beam$^{-1}$ to Jy pix$^{-1}$ by dividing by the number of pixels in the beam.  Following this method a set of Stokes $Q$ and $U$ maps with the resolution, noise and frequencies described by \cite{2023ApJ...955...16B} were created from our numerical model and using RM synthesis with uniform weighting we produced an RM map (see the third panel down on the LHS of Fig. \ref{Baidoo_001}).

\begin{table*}
\caption{Conversion of resolution and Stokes $Q$ and $U$ channel noise to simulation units.  The column headings are:  Run number; Name of object; redshift (z); Luminosity distance (D$_{\text{L}}$); Resolution in arcsecond; Flux of source in Jansky ($F_{\text{S}}$); Stokes $Q$ and $U$ channel noise in milliJansky per beam; Resolution in pixels per beam; Flux of numerical model in simulation units ($F_{\text{M}}$); Stokes $Q$ and $U$ channel noise in simulation units per pixel.  For run 1: Flux value is from \protect\cite{2017ApJS..230....7P} at a viewing frequency of $\num{2e9}$ Hz; redshift, resolution (in arcsecond) and noise (in mJy beam$^{-1}$, midpoint value of published range used) are from \protect\cite{2023ApJ...955...16B}; Luminosity distance and Scale are calculated using the redshift value in the online calculator \protect\url{https://www.astro.ucla.edu/~wright/CosmoCalc.html} and assuming a flat $\Lambda$CDM cosmology with H$_0 = 69.3$ km s$^{-1}$ Mpc$^{-1}$ and $\Omega_{\text{M}} = 0.288$ (from \protect\cite{2023ApJ...955...16B}); Flux of the model is calculated from the simulation; model resolution and noise are then calculated from these values (see text).  For other runs: redshift; resolution (in arcsecond); flux and Stokes $Q$ and $U$ noise in mJy beam$^{-1}$ observed at a frequency of 144 MHz, are taken from from the Virtual Observatory \protect\url{https://dc.g-vo.org} (referenced in \protect\cite{2023MNRAS.519.5723O}).}
\label{noise_values}
\centering
\begin{tabular}{ccccccccccc}
\hline
 Run & Name & z & D$_{\text{L}}$ & Scale & Resolution & $F_{\text{S}}$ & Noise & Resolution & $F_{\text{M}}$ & Noise\\
 & & & (Mpc) & (kpc/") & (") & (Jy) & (mJy/beam) & (pix/beam) & (sim.) & (sim./pix)\\
 \hline
 1 & 3C218 (Hydra A) & 0.054 & 243.3 & 1.062 & 1.5 $\times$ 1.0 & 30 & 0.58 & 1.9 & \num{4.64e-7} & \num{4.77e-12} \\
 2 & ILTJ115858.50+582040.8  & 0.0537 & 222.5 & 1.079 & 20 & 1.5948 & 1.624 & 518 & \num{4.64e-7} & \num{9.11e-13}\\
 3 & ILTJ144156.19+340128.9 & 0.0871 & 410.2 & 1.683 & 20 & 0.1722 & 3.276 & 1261 & \num{4.64e-7} & \num{6.99e-12}\\
 4 & ILTJ142344.98+633939.8 & 0.204 & 1033.3 & 3.456 & 20 & 3.8374 & 1.816 & 5318 & \num{4.64e-7} & \num{4.12e-14}\\
 \hline
\end{tabular}
\end{table*}

The RM map generated from the RM synthesis technique can be seen on the third panel down on the left hand side of Fig. \ref{Baidoo_001}; this can be compared with the map directly from the simulation (top panel) or the simulated map smoothed with a Gaussian to the same resolution of observations (second panel down); the mean absolute difference between the top and third panels can be seen in the bottom panel.  In addition, the right hand panel of Fig. \ref{Baidoo_001} compares the Faraday dispersion function from RM synthesis (in black) with the RM value direct from the simulation (red) for various identified lines of sight (LOS) in the RM maps (points A to F).  It can be seen that RM synthesis is successful in recreating the RM map; the peak of the Faraday dispersion function coincides almost exactly with the value from the simulation for the LOS identified and the overall pattern of the RM map is faithfully replicated.  The only area of discrepancy is that values at the very edge of the lobes are less successfully reproduced as a result of the noise added to our $Q$ and $U$ maps before using RM synthesis.  The values of Stokes $Q$ and $U$ are very low at the edges of the lobe and so the resultant signal tends to be drowned out by the noise near the edge.  If these were genuine observations then such values would be removed by the image processing algorithms.  We conclude that we have been able to use RM synthesis to obtain the RM map from our simulated AGN when we set the observational distance, resolution and noise level to similar values of current observations of Hydra A as presented in \cite{2023ApJ...955...16B}.

\subsection{RM Synthesis of LOFAR Observations}
We use our model simulation as a representation of LOFAR objects by setting the resolution and noise level of our model to match those seen in observations.  Using the RM Grid of \cite{2023MNRAS.519.5723O} and in particular data from the Virtual Observatory \url{https://dc.g-vo.org}, we find data for three example LOFAR objects (see table \ref{noise_values}).  We then follow the same method as described in Section \ref{Baidoo_RMsynthesis} in order to calculate model resolution and Stokes $Q$ and $U$ noise (equation \ref{noise}) and then to obtain our RM maps using RM synthesis.  In order to ensure that our values are as realistic as possible, we restrict our study to LOFAR objects with a similar linear size as our model (i.e. $\sim$500 kpc).

\begin{figure*}
\begin{center}
\begin{minipage}[c]{0.6\linewidth}
\includegraphics[width=1\textwidth]{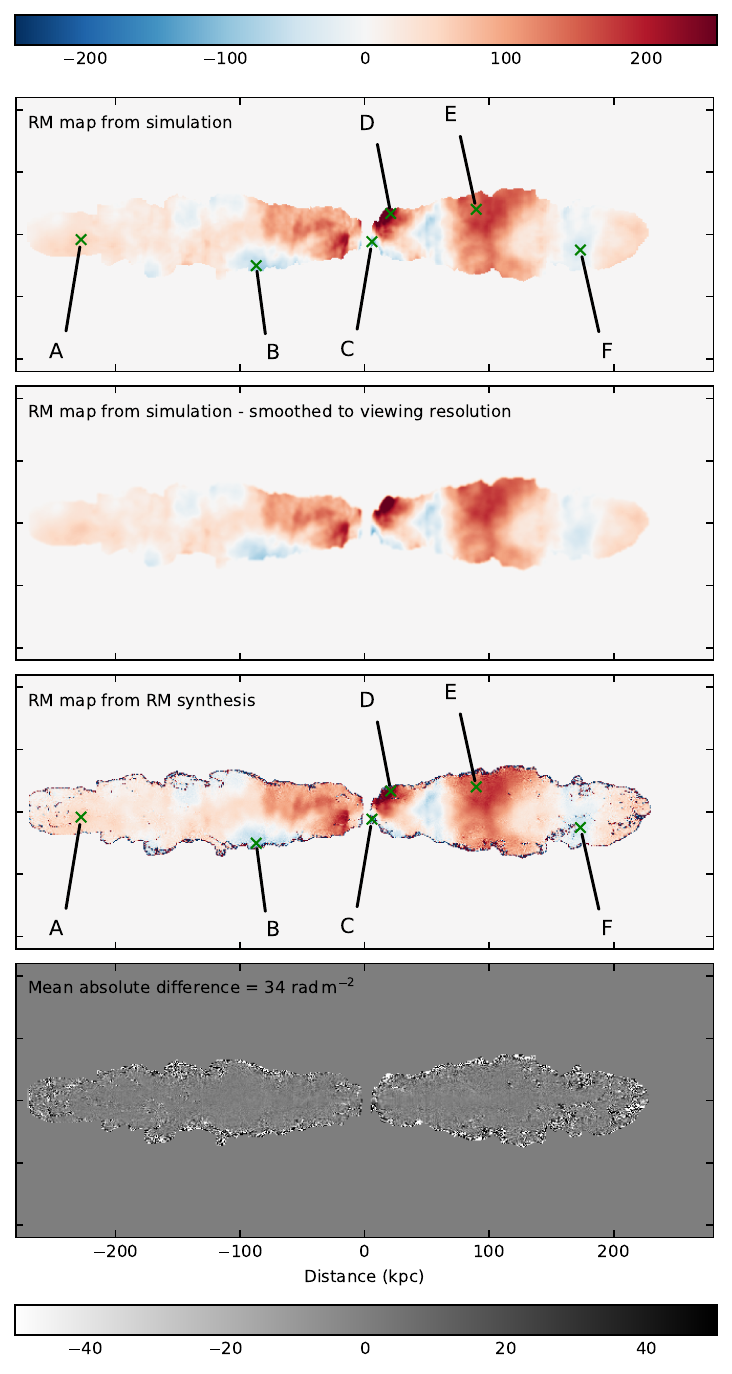}
\end{minipage}
\hfill
\begin{minipage}[c]{0.38\linewidth}
\includegraphics[width=1\textwidth]{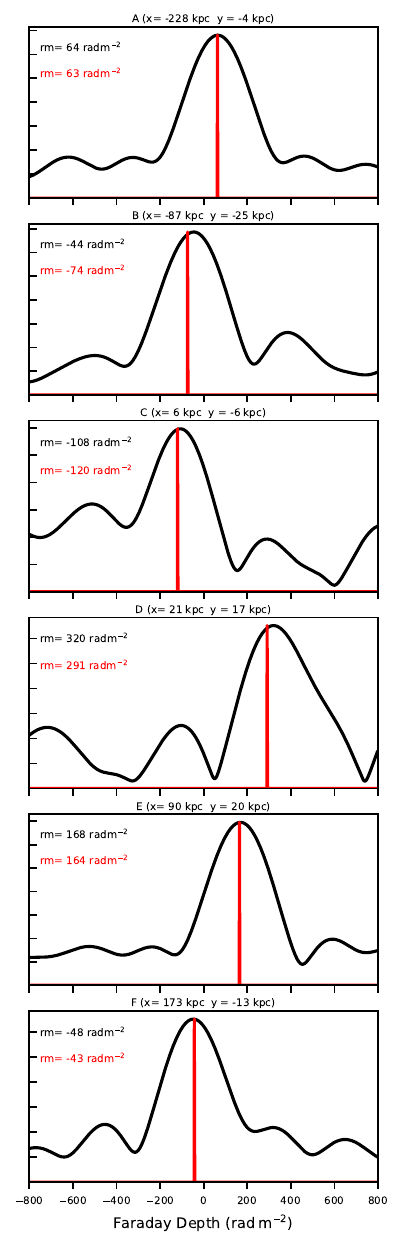}
\end{minipage}
\end{center}
\caption{RM maps from a simulation and from RM synthesis based upon the observations of 3C218 (Hydra A) by \protect\cite{2023ApJ...955...16B}.  The left panel (from the top) shows: RM map directly from the Faraday screen of our simulation; simulation RM map convolved with a Gaussian to give a resolution similar to that of the observations; RM map from RM synthesis (derived from the frequency distribution, resolution and Stokes $Q$ and $U$ channel noise as described in \protect\cite{2023ApJ...955...16B}, see table \ref{noise_values}); map of mean absolute difference between the RM map of the simulation (top) and that created by RM synthesis.  The green crosses show the lines of sight (LOS) corresponding to the right panel showing RM values from the simulation (in red) and the Faraday dispersion function calculated from RM synthesis (in black, peak value also in black).  The model is for a jet of power $\num{1e38}$W running into an atmosphere of $M_{500}=3\times\num{e14}\,\text{h}^{-1}_{70}M_{\odot}$. RM values are for external Faraday rotation only.  The y-axes of the Faraday dispersion function charts increase in steps of $0.001$.}
\label{Baidoo_001}
\end{figure*}

\begin{figure*}
\begin{center}
\begin{minipage}[c]{0.6\linewidth}
\includegraphics[width=1\textwidth]{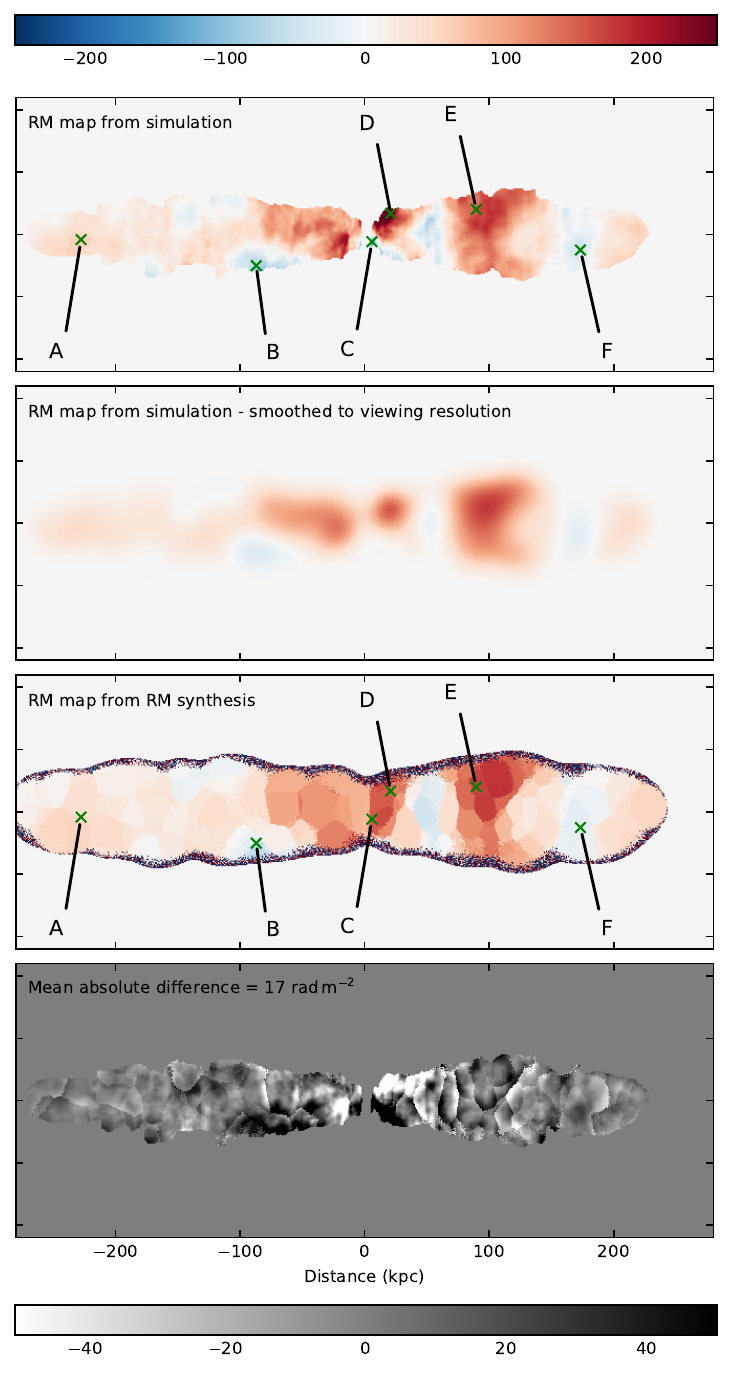}
\end{minipage}
\hfill
\begin{minipage}[c]{0.38\linewidth}
\includegraphics[width=1\textwidth]{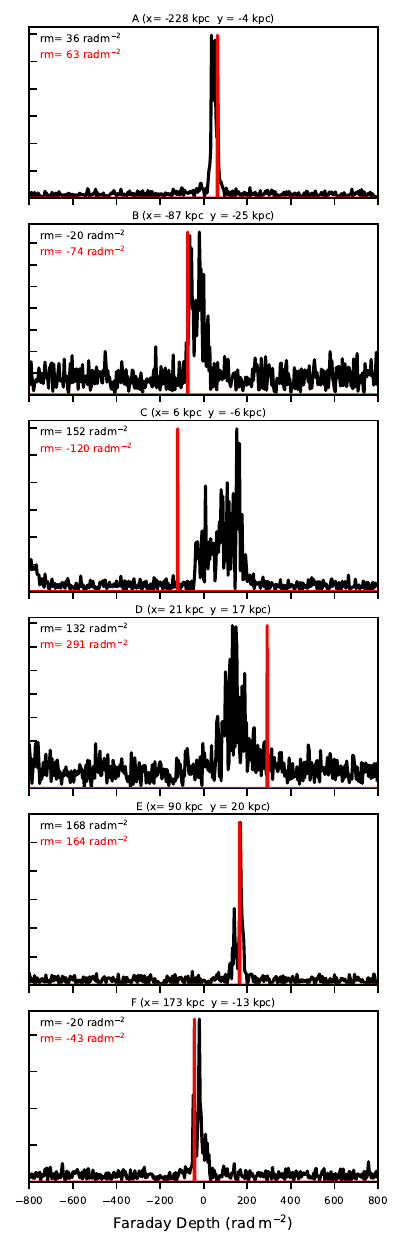}
\end{minipage}
\end{center}
\caption{Same as Fig. \ref{Baidoo_001} except based upon the LOFAR observations of ILTJ115858.50+582040.8 from the RM Grid by \protect\cite{2023MNRAS.519.5723O} (and data from the Virtual Observatory \url{https://dc.g-vo.org}).  Corresponds to run 2 of table \ref{noise_values}.}
\label{LOFAR_010}
\end{figure*}

\begin{figure*}
\begin{center}
\begin{minipage}[c]{0.6\linewidth}
\includegraphics[width=1\textwidth]{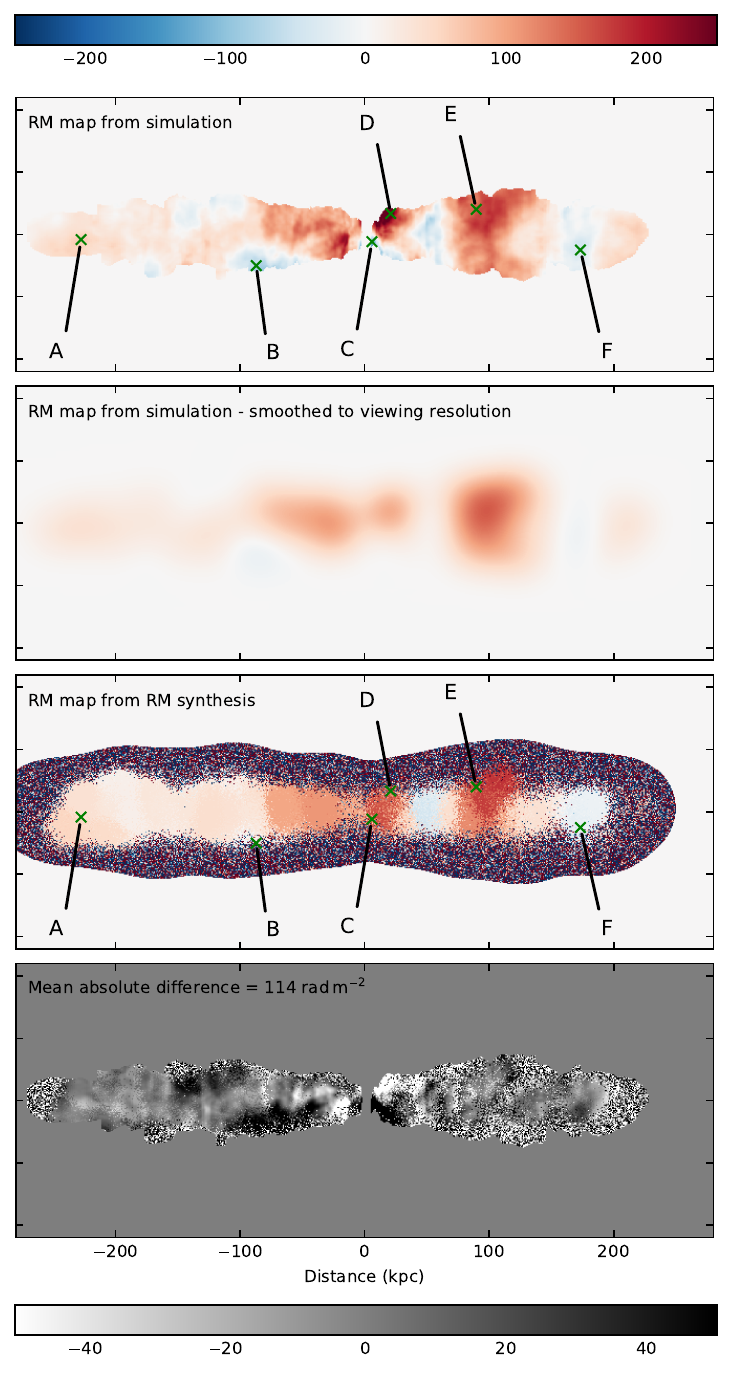}
\end{minipage}
\hfill
\begin{minipage}[c]{0.38\linewidth}
\includegraphics[width=1\textwidth]{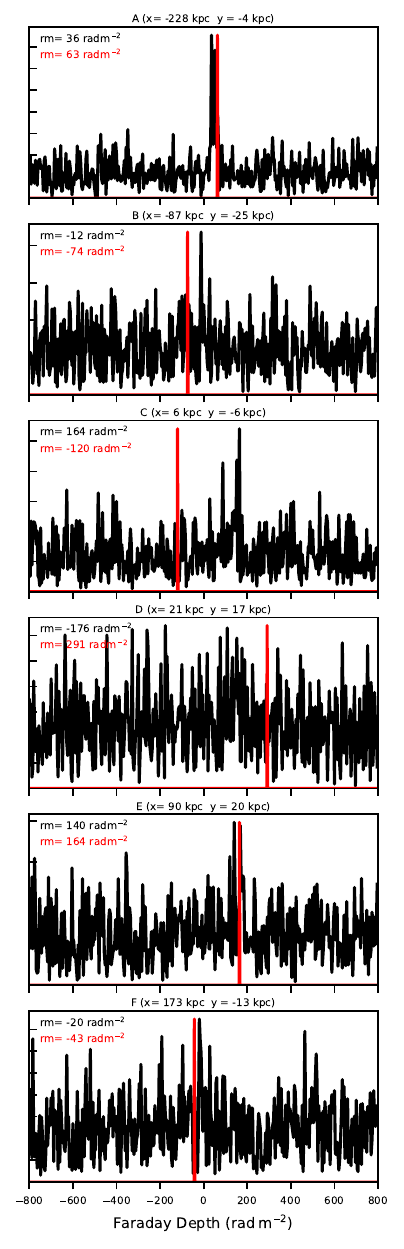}
\end{minipage}
\end{center}
\caption{Same as Fig. \ref{Baidoo_001} except based upon the LOFAR observations of ILTJ144156.19+340128.9 from the RM Grid by \protect\cite{2023MNRAS.519.5723O} (and data from the Virtual Observatory \url{https://dc.g-vo.org}).  Corresponds to run 3 of table \ref{noise_values}.}
\label{LOFAR_016}
\end{figure*}

\begin{figure*}
\begin{center}
\begin{minipage}[c]{0.6\linewidth}
\includegraphics[width=1\textwidth]{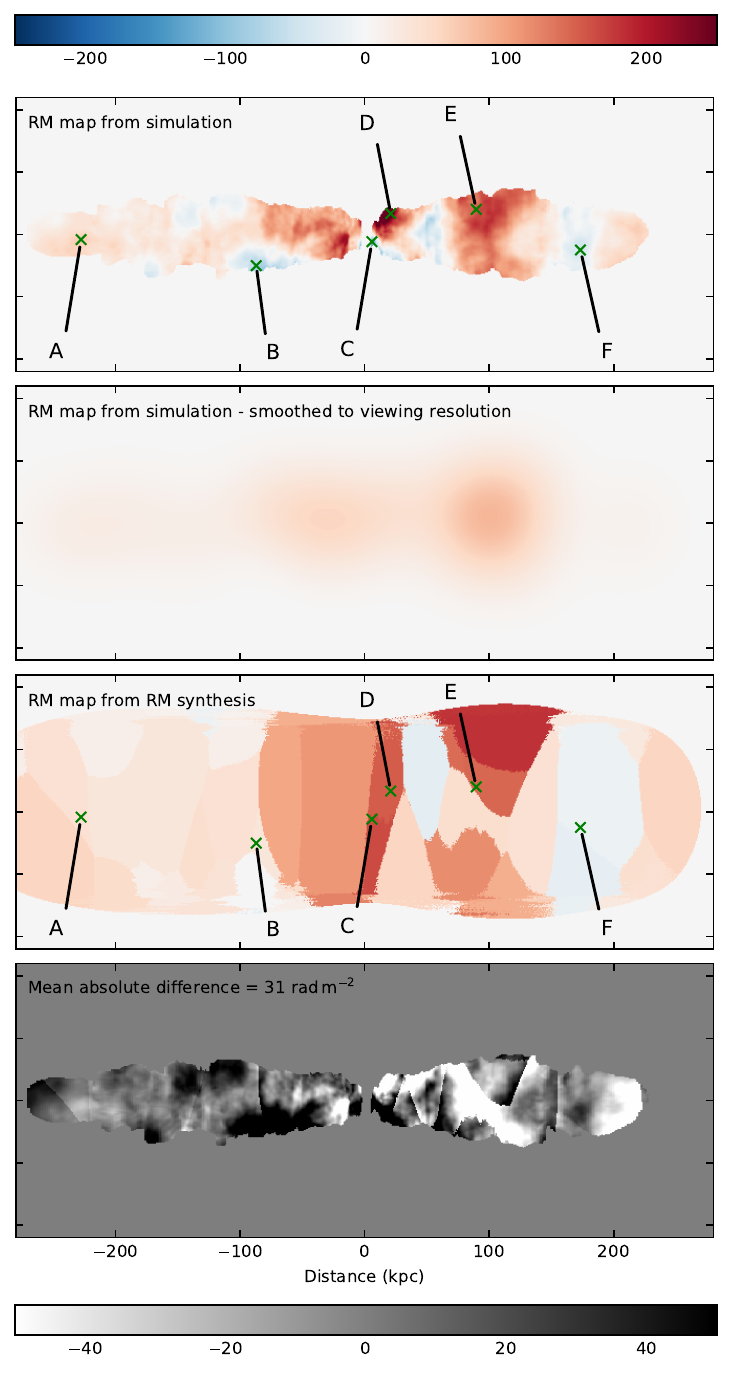}
\end{minipage}
\hfill
\begin{minipage}[c]{0.38\linewidth}
\includegraphics[width=1\textwidth]{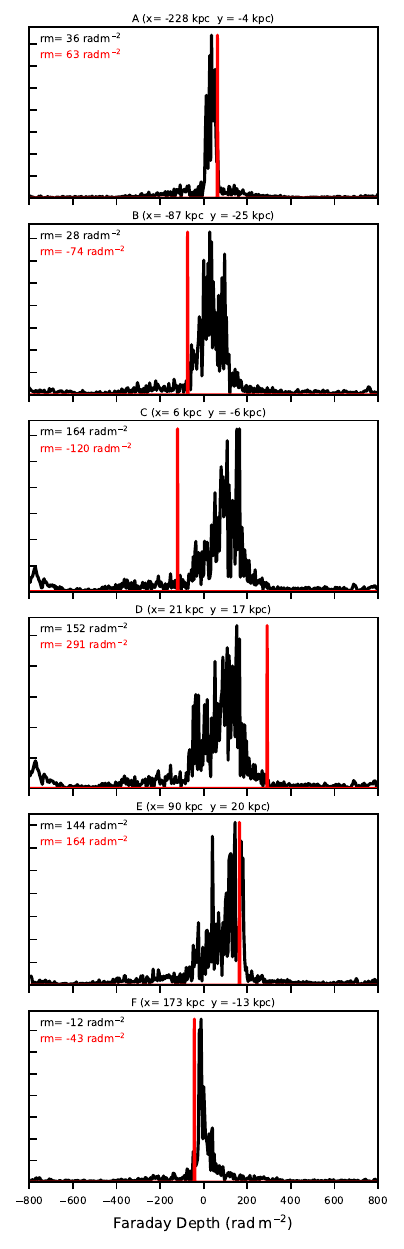}
\end{minipage}
\end{center}
\caption{Same as Fig. \ref{Baidoo_001} except based upon the LOFAR observations of ILTJ142344.98+633939.8 from the RM Grid by \protect\cite{2023MNRAS.519.5723O} (and data from the Virtual Observatory \url{https://dc.g-vo.org}).  Corresponds to run 4 of table \ref{noise_values}.}
\label{LOFAR_033}
\end{figure*}

The same analysis was carried out as above for these LOFAR objects, the charts produced can be seen in Figs \ref{LOFAR_010}, \ref{LOFAR_016} and \ref{LOFAR_033}.  RM synthesis has less success in recreating the RM map in comparison with the Hydra A example in Section \ref{Baidoo_RMsynthesis}; this is as a result of the lower resolution, lower frequency and higher noise involved in LOFAR observations.  In Fig. \ref{LOFAR_016} the impact of lower signal-to-noise ratio can be seen in the RM map as a `speckled' pattern around the edges, as well as on the right panel where the Faraday dispersion function has many spikes; the peak value is not clearly defined and RM synthesis often chooses a different value to the `correct' value direct from the simulation.  The impact of a lower resolution object (i.e. an object which is further away) can be seen in Fig. \ref{LOFAR_033} where, as well as spreading out over a larger area, the RM map forms a `patchwork' of constant values.  This `patchwork' is also seen in Fig. \ref{LOFAR_010}.  The RM maps from the simulation show a gradual variation in RM across the map without regions of constant value; nevertheless, the `patches' seen in the maps generated by RM synthesis do successfully produce the same rough pattern and values, albeit with reduced precision.  These LOFAR examples, whilst resulting in RM maps of lower quality than the higher resolution example in Fig. \ref{Baidoo_001}, nevertheless do produce results which reflect the original RM map from the simulation, while the faithfulness of these to the true RM map is clearly limited by resolution and noise effects.



\subsection{RM Synthesis: RM between lobes}\label{RMsynth_test}
\begin{figure*}
\centering
\begin{minipage}{.49\textwidth}
\begin{subfigure}{\textwidth}
\centering
\includegraphics[width=1.0\textwidth]{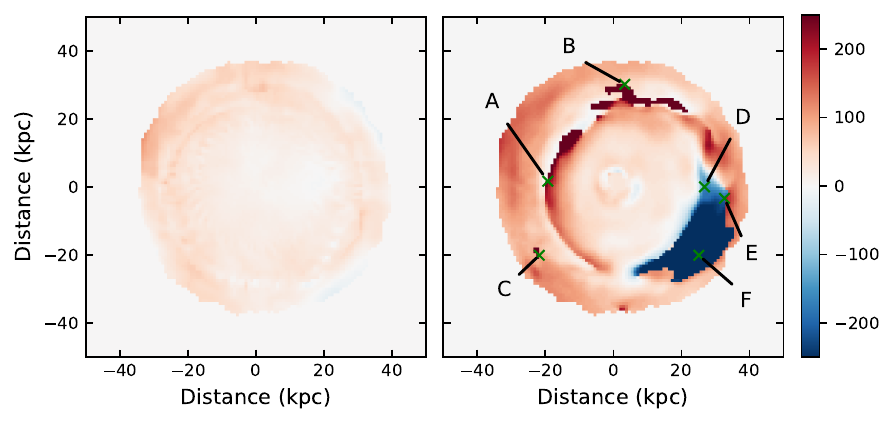}
\end{subfigure}\\
\begin{subfigure}{\textwidth}
\centering
\includegraphics[width=1.0\textwidth]{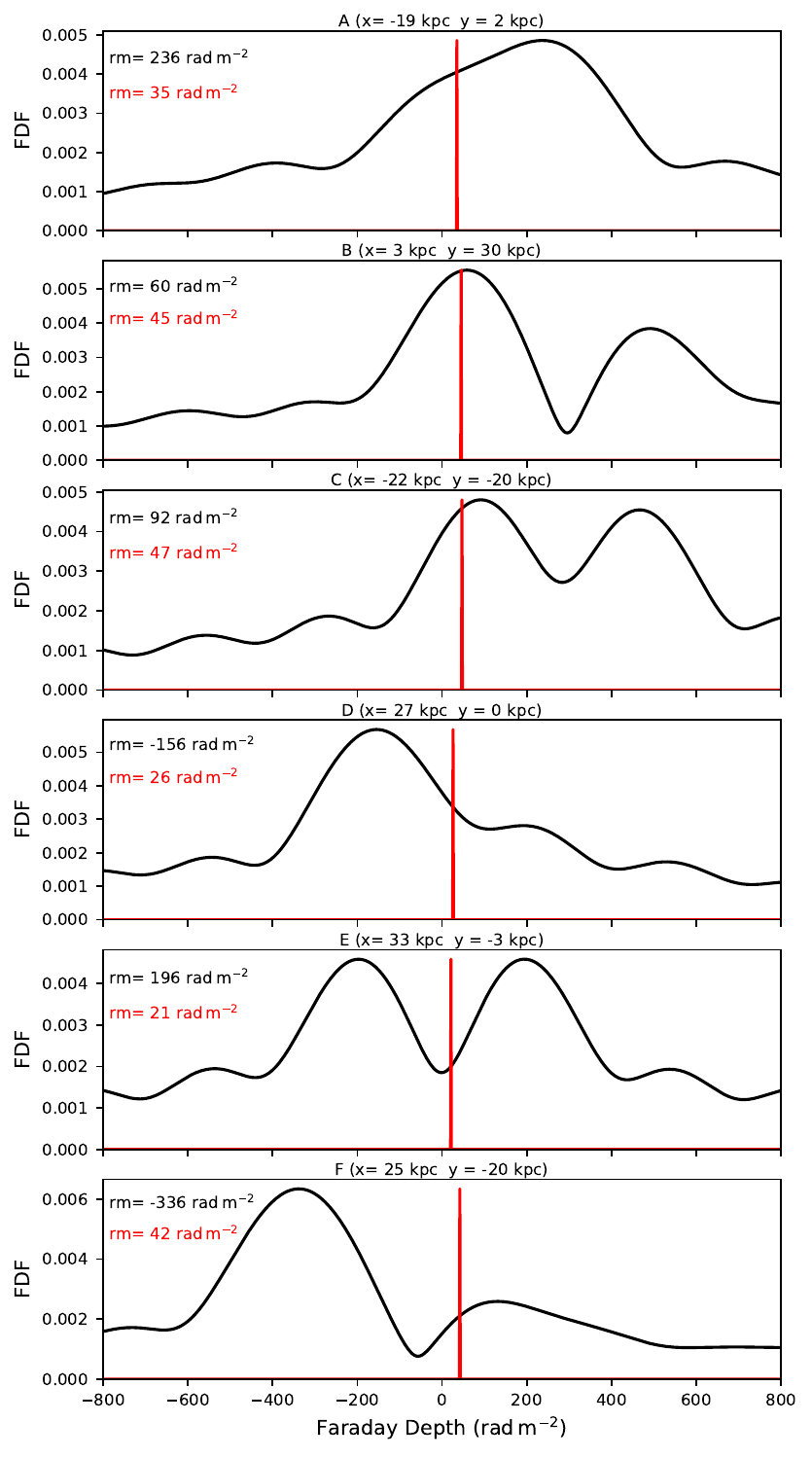}
\end{subfigure}%
\end{minipage}
\hfill
\begin{minipage}{.485\textwidth}
\begin{subfigure}{\textwidth}
\centering
\includegraphics[width=1.0\textwidth]{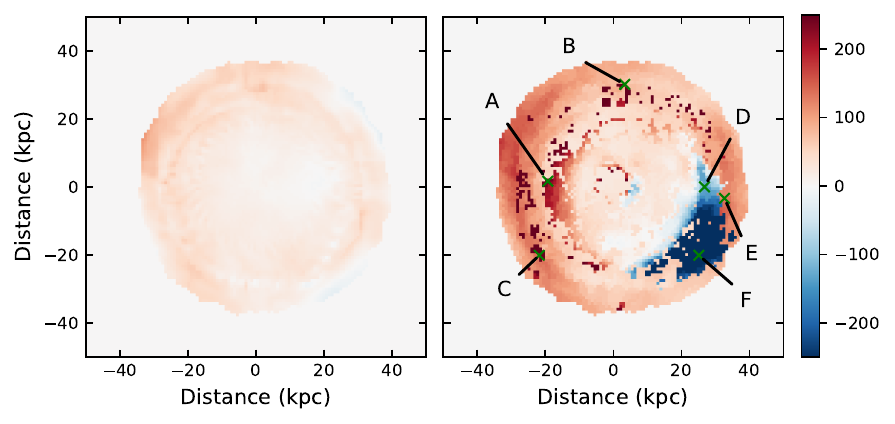}
\end{subfigure}\\
\begin{subfigure}{\textwidth}
\centering
\includegraphics[width=1.0\textwidth]{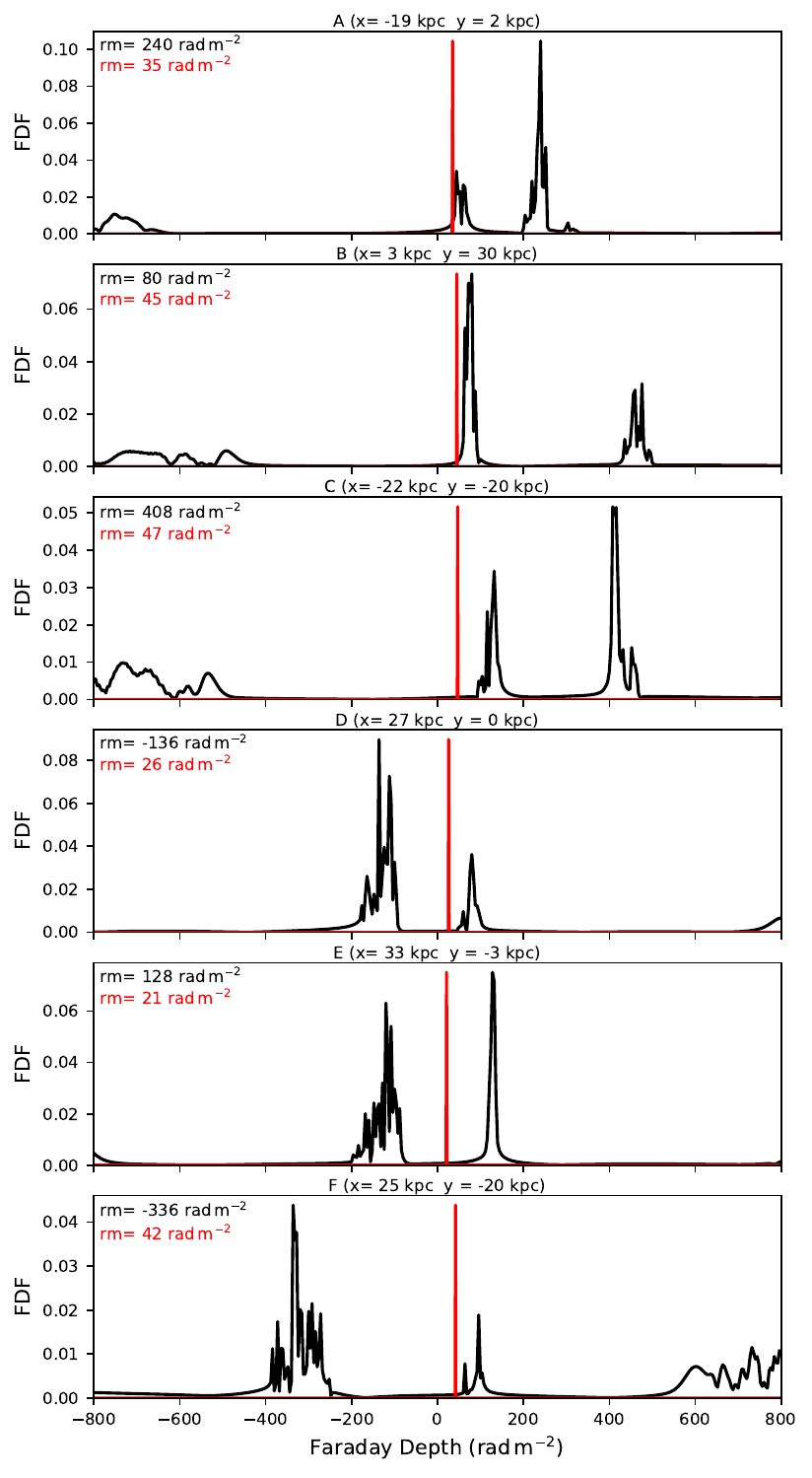}
\end{subfigure}%
\end{minipage}%
\caption{Ability of RM synthesis to distinguish two or more sources - observation along jet axis.  Top left panel:  RM maps for external Faraday screen only (left) and for all material inside and between lobes (right), with labelled lines of sight (LOS).  Bottom left panel:  Faraday spectra of LOS from RM synthesis (black) and from the simulated external Faraday screen (red).  Results here are for a jet of power $\num{1e38}$W running into an atmosphere of $M_{500}=3\times\num{e14}\,\text{h}^{-1}_{70}M_{\odot}$; are derived from the frequency distribution as described in \protect\cite{2023ApJ...955...16B}.  Top right and bottom right, same as LHS but for the LOFAR frequency distribution as described by \protect\cite{2021MNRAS.502..273M}.}
\label{FD_alongjet}
\end{figure*}

A major advantage of the RM synthesis technique is its ability to distinguish between two or more sources of RM along a single line of sight.  As a test of our implementation of the RM synthesis method we examine here the capacity to discern more than one source of RM along a single LOS.  We do this by looking along the jet axis of our model and consider both external and internal contributors to the Faraday dispersion.  We use the fiducial model of a jet of power $\num{1e38}$W running into an atmosphere of $M_{500}=3\times\num{e14}\,\text{h}^{-1}_{70}M_{\odot}$, once it has reached an average lobe length of $250$ kpc.  We obtain our RM map by ray-tracing from the shock-lobe boundary at the most distant part of the more distant lobe (inflated by the counterjet), towards us through that lobe, through the intervening cluster material between the two lobes, through the lobe created by the jet closest to us and finally through the material between that nearest lobe and the observer.  As the LOS passes towards us through lobe material, synchrotron emission from each cell contributes to the Stokes parameters, in addition, each cell (lobe and outside the lobe) causes Faraday rotation (although our study has shown that, for the lobe densities used here, this is very small for lobe material).  The top panels of Fig. \ref{FD_alongjet} show the RM map of just the external Faraday screen (left) and also the sum total of the RM through both lobes and intervening cluster material (right); where it is clear that including contributions through the entire model leads to Faraday depth values that are significantly greater/different to RM derived just from the external Faraday screen.  The lower panel displays the Faraday dispersion function for labelled lines of sight through the model.  It is clear that there are two main sources presented in these Faraday spectra: one corresponds to the Faraday screen between the nearest lobe and the observer (with internal depolarization from the nearest lobe); while the other is from the cluster material between the lobes.  The conclusion we draw here is that our RM synthesis technique is able to disentangle the signal and reveal two (or more) sources, as expected.

\section{SUMMARY AND CONCLUSIONS}
In this study we have derived polarization products from our numerical model of a spherically symmetric cluster atmosphere based upon the Universal Pressure Profile; the most realistic general atmosphere yet.  We simulated some of the polarimetric observations of Hydra A (in particular the work of \cite{2023ApJ...955...16B}) and a selection of high-RM LOFAR objects.  We produced fractional polarization maps which demonstrate the observational trends of a decrease in polarization with frequency and resolution; these results reveal the same trends as recent observations.  We also produced depolarization maps which, as well as being very similar to those of \cite{2023ApJ...955...16B}, are able to demonstrate that we placed our AGN at the centre of our cluster as the measure of depolarization traces out the density profile of our models.  We conclude that depolarization can be used as a proxy for the cluster density profile; although we must be mindful that, as highlighted by \cite{2024MNRAS.531.2532J}, the RM map depends critically upon the shape of the lobe and so this also has a great influence upon the amount of depolarization.

We reproduced some of the results we saw in our previous work on polarimetric simulations, such as the Laing-Garrington effect.  As well as demonstrating the effect, we varied the angle of view in order to demonstrate that the effect is greater for angles closer to the jet-counterjet axis, as expected as the RM would be greater for the more distant lobe.  If magnetic field structures in real clusters are similar to the multi-scale turbulent one that we have used here, the accuracy with which the orientation of the jet axis to the line of sight can be determined is limited to tens of degrees at best.  In addition, our fractional polarization histograms, as well as reproducing the expected trends in polarization with frequency and resolution, were more faithful to those seen in observations than in our previous work.  This is likely as a result of improvements in our numerical model as we used a smaller (and more realistic) injection cylinder as well as a helical injected magnetic field in our jet.

In conducting a study of the variation of Mean Fractional Polarization (MFP) with frequency for a range of resolutions, we found that results were in agreement with expectations (lower MFP with lower resolution).  Furthermore, our results follow the general trend of the Burn law \citep{1966MNRAS.133...67B} in describing how the MFP decreases below a certain frequency.  

We investigated the variation of MFP with frequency (for a fixed resolution) for the full range of cluster mass and jet power of our models and conclude that both cluster mass and jet power influence the variation of MFP.  We found that MFP decreases with increasing cluster mass and increasing jet power.  These results can be explained by the RM properties of the Faraday screen:  We calculated the mean absolute RM for each of our runs and demonstrated that greater cluster mass and greater power jets have larger values of RM.  The impact of clusters of greater mass is accounted for by larger values of density and magnetic field strength (so RM is greater) but the impact of jet power is influenced by the morphology of the resultant lobes; whereby more of the higher power lobes exist behind regions of higher RM (i.e. near the core).  We identified from physical principles a critical frequency for internal depolarization (the frequency below which depolarization becomes important) which supported our trend of increased cluster mass leading to increased depolarization.

Rotation measure was visualised in a number of ways; one way was to view our model from different angles and so observe the variation of RM for a sequence of images.  This enabled us to see that values for RM were considerably higher when viewing the more distant lobe at a smaller angle to the jet-counterjet axis.  This enables us to visualise the reason behind the Laing-Garrington effect; a differential in RM between the lobes when the AGN is not aligned with the plane of the sky.

We used the RM synthesis technique to create RM maps from simulated Stokes $Q$ and $U$ images, which were adjusted to reflect the resolution and noise of real observations of Hydra A and LOFAR.  We found that, whilst the process is successful in recreating the simulated RM map, this can be limited by both the resolution and the noise level of the source.  The lower resolution results in a patchwork of RM values as the Faraday dispersion function (FDF) is spread over a larger range of Faraday depth; whereas the increased noise level is seen on the FDF as many spikes.  It was seen that the noise impacts more those LOS with lower values of Stokes $Q$ and $U$ (i.e. near the edges of the lobe, where the LOS passes through less synchrotron emitting material as the lobe is thinnest here) leading to greater uncertainty of RM value here.  In real observations such values would be removed as they would fall below a pre-determined signal-to-noise ratio.

In this study we have utilised LOFAR objects identified within the RM Grid of \cite{2023MNRAS.519.5723O}; therefore, our examples have detectable RM values and so we do not present `typical' LOFAR objects but only those whose RM screens have been detected by current instrumentation.  Even within the examples we have provided here, it is clear that RM synthesis has difficulty in recovering the RM map of many objects, especially at low frequencies, and only for those examples which are bright and nearby are we able to obtain a detailed RM map using RM synthesis.  A key limitation of LOFAR is the low resolution used to create the RM Grid (i.e. 20") and in future, should higher resolutions be used, then this will enable the recovery of the RM map from RM synthesis for more distant and less powerful examples.

We used RM synthesis to successfully demonstrate the expected result of looking down along the jet axis of our simulation through both lobes; we identified multiple components.   This important test highlights the fact that a normal straight non-restarting jet can be identified by RM synthesis as a single RM source, provided it is in the plane of the sky, whereas more complicated scenarios (such as looking end-on, restarting jets, precessing jets) have more complicated RM synthesis results which evidence evolutionary histories.  One example is Hydra A, the RM synthesis results in \cite{2023ApJ...955...16B} demonstrate that along some lines of sight there are additional Faraday screens most likely as a result of entrained gas, possibly due to the jet re-starting or changing direction.  A future scope of work would be to synthesise such evolutionary activity within our models and find links to the characteristic RM synthesis output; this would prove a very useful tool for RM synthesis observations of AGN.  Our study has shown that this work would certainly be possible for VLA objects, and also for bright, nearby LOFAR objects.

\section*{Acknowledgements}

This work has made use of the University of Hertfordshire Science and Technology Research Institute high performance computing facility (\url{https://uhhpc.herts.ac.uk/}).  We also thank an anonymous referee for helpful comments which led to improvements upon the original text.

\section*{Data Availability}

The data underlying this article will be shared on reasonable request to the corresponding author.



\bibliographystyle{mnras}
\bibliography{References} 

\appendix


\section{Maximum value for fractional polarization}\label{100_frac}
We argue that synchrotron emission from aged sources can have a fractional polarization higher than the widely accepted value of $\sim 0.7$, and potentially up to $100$ per cent.  In our context, we are dealing with the synchrotron emission from large scale regions (i.e. AGN lobes) which would be expected to have an aged electron spectrum which steepens from an initial power law before cutting off completely at some critical energy.  Because of the different behaviour of total and polarized emission, such an electron population would be expected to have a frequency-dependent fractional polarization, which increases as the spectrum steepens, and can approach 100 per cent as the exponential cutoff is reached.

The polarized emissivity is calculated in the standard way by integrating over the electron population and pitch angle distribution,
\begin{equation}
J_p(\nu)=\int^{\pi}_0\int^{E_{\text{max}}}_{E_{\text{min}}}\frac{\sqrt{3}Be^3\sin{\alpha}}{16\pi^2\epsilon_0cm_e}G(x)N(E)\sin{\alpha}\mathrm{d}E\mathrm{d}\alpha
\end{equation}
where $x$ is a dimensionless function of the frequency, field strength and energy:
\begin{equation}
x=\frac{\nu}{\nu_c}=\frac{4\pi m^3_ec^4\nu}{3eE^2B\sin{\alpha}}
\end{equation}
and G(x) is the conventional way \citep{1986rpa..book.....R} of expressing the dimensionless frequency dependence of single-electron polarized emissivity:
\begin{equation}
G(x)=xK_{2/3}(x)
\end{equation}
with $K_{2/3}$ the modified Bessel function of order 2/3.

\begin{figure*}
\begin{center}
\includegraphics[width=0.8\textwidth]{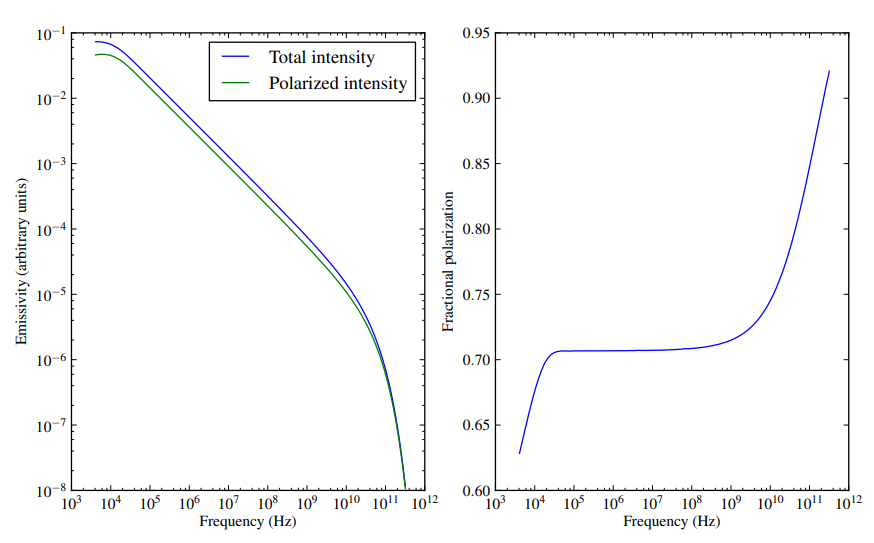}
\end{center}
\caption{Left: total and polarized intensity as a function of frequency for an aged electron population.  Right: fractional polarization as a function of frequency.  The example spectrum used has $\gamma_{\text{min}}=10$ and $p=2.2$; the mean magnetic field is $1.5$ nT, and the spectral age $\num{3e6}$ years.  The fractional polarization is consistent with the standard power-law calculation (see text) in the region in which the total-intensity spectrum is a power law, but deviates in the regions of low- and high-frequency cutoffs.}
\label{Hardcastle_steepspectrum}
\end{figure*}

When intrinsic polarization of synchrotron radiation is discussed it is often assumed that the electron energy distribution $N(E)$ is a power law $N_0E^{-p}$; in this case it is well known \citep{1986rpa..book.....R} that the ratio of emissivities can be shown to be independent of frequency, giving a fractional polarization for $p = 0.5$: $\mu=(p +1)/(p +5/3)=0.69$.  However, we expect large-scale regions of synchrotron emission in radio galaxies to have an aged electron spectrum which steepens from an initial power law before cutting off altogether at some critical energy; here we use the Jaffe-Perola model \citep{1973A&A....26..423J}.  Because of the different behaviour of polarized and total emissivity, such an electron population will have a frequency-dependent fractional polarization, which increases as the spectrum steepens, and can approach $100$ per cent as the exponential cutoff is reached (see Fig. \ref{Hardcastle_steepspectrum}).

\bsp	
\label{lastpage}
\end{document}